\documentclass[aps,pra,twocolumn,showpacs,amsmath,amssymb,nofootinbib,longbibliography,superscriptaddress
]{revtex4-1}

\usepackage[pdftex]{graphicx}
\usepackage[usenames,dvipsnames]{xcolor}
\usepackage{epstopdf}
\usepackage{tikz}
\usepackage{mathtools}

\usepackage[normalem]{ulem}

\graphicspath{{FIGs/}}

 \usepackage[utf8]{inputenc}
\usepackage[T1]{fontenc}

\usepackage{beramono}
\usepackage{listings}

\usepackage{lmodern}

\usepackage{amssymb,amsmath,amsthm}

\usepackage{subcaption}

\usepackage{algorithm}
\usepackage{algpseudocode}

\usepackage[justification=raggedright,singlelinecheck=false]{caption}

\theoremstyle{definition}

\theoremstyle{remark}

\makeatletter
\DeclareFontEncoding{LS1}{}{}
\DeclareFontEncoding{LS2}{}{\noaccents@}
\DeclareFontSubstitution{LS1}{stix}{m}{n}
\DeclareFontSubstitution{LS2}{stix}{m}{n}
\makeatother

\DeclareMathAlphabet\mathscr{LS1}{stixscr}{m}{n}
\SetMathAlphabet\mathscr{bold}{LS1}{stixscr}{b}{n}

\makeatletter
\lst@InstallKeywords k{attributes}{attributestyle}\slshape{attributestyle}{}ld
\lst@InstallKeywords k{attributes2}{attributestyle2}\slshape{attributestyle2}{}ld
\lst@InstallKeywords k{attributes3}{attributestyle3}\slshape{attributestyle3}{}ld
\lst@InstallKeywords k{attributes4}{attributestyle4}\slshape{attributestyle4}{}ld 
\lst@InstallKeywords k{attributes5}{attributestyle5}\slshape{attributestyle5}{}ld 
\lst@InstallKeywords k{attributes6}{attributestyle6}\slshape{attributestyle6}{}ld 
\lst@InstallKeywords k{attributes7}{attributestyle7}\slshape{attributestyle7}{}ld 
\makeatother

\definecolor{macros}{rgb}{0.64,0.55,0.08}
\definecolor{types}{rgb}{0,0.55,0.55}
\definecolor{functions}{rgb}{1,0.27,0}

\definecolor{juliaterminal}{rgb}{0,0.79,0}
\definecolor{juliapkg}{rgb}{0.49,0,1}
\definecolor{juliahelp}{rgb}{0.59,0.59,0}
\definecolor{juliamodules}{rgb}{0,0.79,0.79}

\lstdefinelanguage{julia}%
  {morekeywords={abstract,break,case,catch,const,continue,do,else,elseif,%
      end,export,false,for,function,immutable,import,importall,if,in,nothing,%
      macro,module,otherwise,quote,return,switch,true,try,type,typealias,%
      using,while,mutable},%
   sensitive=true,%
   alsoother={\$},
   morecomment=[l]\#,%
   morecomment=[n]{\#=}{=\#},%
   morestring=[s]{"}{"},%
   morestring=[m]{'}{'},%
   moreattributes={@inline,@inbounds,@simd,@threads},
   attributestyle = \bfseries\color{macros}, 
moreattributes2={AbstractArray,Array,Union,Matrix,Vector,Number,Integer,Tuple,NTuple,Bool,Colon,struct,Real,Float64,ComplexF64,Int64,include,UnitRange,DataType,Function,String,StepRange,undef,TensorPACK},
   attributestyle2 = \bfseries\color{types}, 
moreattributes3={zeros,size,eigen,contract,ccontract,contractc,ccontractc,TensType,MPS,MPO,tens,Env,,intType,svd,unreshape,denstens,qarray,qr,lq,move,spinOps,makeMPO,makeMPS,expect,largeMPO,largeMPS,largeEnv,fermionOps,spinOps,tJOps,environment,makeEnv,applyMPO,largematrixproductstate,largematrixproductoperator,largeenvironment,correlation,correlationmatrix,polar,matrixproductoperator,matrixproductstate,loadMPS,loadMPO,loadLenv,loadRenv,transfermatrix,dmrg,Lupdate,Rupdate,boundaryMove,div,lanczos,twosite_update,twositeOps,simpledmrg,fullH,randMPS,applyOps,applyOps,fullpsi,sub,div,add,mult,tenstype,intvecType,DMRjulia,genColType,makeId,makedens,convertTens,makeArray,convIn,findnotcons,checkType,tensorcombination,inverse_element,invmat,showTens,makepos,position_incrementer,get_ranges,pos2ind,ind2pos,maincontractor,corecontractor,prepareT,permutedims_2matrix,matrixequiv,libqr,liblq,regEnv,vecenvironment,makeoc,elnumtype,getSingleVal,joinindex,libmult,trace,checkcontract,envType,regMPS,regMPO,regEnv,tensor2disc,tensorfromdisc,compute_alpha,lanczos,compute_beta,makeEnds,densTensType,findnewm,safesvd,recursive_SVD,libsvd,libeigen,libqr,liblq,largeType,operator_in_order,moveL,moveR,movecenter,leftnormalize,rightnormalize,permutations,largeLenv,largeRenv,largeEnv,loadEnv,penalty},   
   attributestyle3 = \bfseries\color{functions},
moreattributes4={julia},
   attributestyle4 = \bfseries\color{juliaterminal}, 
moreattributes5={pkg},
   attributestyle5 = \bfseries\color{juliapkg}, 
moreattributes6={help},
   attributestyle6 = \bfseries\color{juliahelp}, 
moreattributes7={Threads,Base,LinearAlgebra,Printf,Serialization,Distributed,BLAS},
   attributestyle7 = \bfseries\color{juliamodules}, 
}[keywords,comments,strings]%

\usepackage[hidelinks]{hyperref}

\begin{document}

\title{Basic linear algebra methods for quantum problems}
\author{Aaron Dayton}
\affiliation{Department of Physics \& Astronomy, University of Victoria, Victoria, British Columbia V8P 5C2, Canada}
\author{Kiana Gallagher}
\affiliation{Department of Physics \& Astronomy, University of Victoria, Victoria, British Columbia V8P 5C2, Canada}
\author{Sarah E.~Huber}
\affiliation{Research Computing Services, University of Victoria, Victoria, British Columbia V8P 5C2, Canada}
\author{Thomas E.~Baker}
\affiliation{Department of Physics \& Astronomy, University of Victoria, Victoria, British Columbia V8P 5C2, Canada}
\affiliation{Department of Chemistry, University of Victoria, Victoria, British Columbia V8P 5C2, Canada}

\begin{abstract}

Making new methods for quantum problems often relies on using basic operations in linear algebra. Often these routines are hidden behind well-known libraries that have been optimized over decades. Attempting to improve on those basic routines would be highly time-consuming. We aim in this article to review those basic routines and provide a knowledge foundation for how to perform basic operations on a computer that would be inaccessible with pen and paper. Elementary details on the solutions to linear algebra problems and computational complexity are reviewed. The focus is on solving eigenvalue problems for quantum systems, but the discussion is generic to many other applications. Common matrix forms relevant to quantum systems and their solution strategies are covered. The discussion extends to computational numerical methods for which the most efficient functions exist in freely available libraries. These include eigenvalue, Schur, QR, LU, LDL, Cholesky, and singular value decompositions. The algorithms for obtaining some of these decompositions are discussed, with focus being placed on those used in modern libraries. 

\end{abstract}
\maketitle

\section{Motivation}

Many problems in mathematics, physical science, and engineering require the solution to a linear equation of the form
\begin{equation}\label{genlineq}
\hat A \mathbf{x} = \mathbf{b}.
\end{equation}
Given that matrix $\hat A$ has dimensions $M \times N$, vector $\mathbf{x}$ must be of length $N$ and vector $\mathbf{b}$ must be of length $M$. For our purposes, it is sufficient to demonstrate basic solution strategies on the square case where $M=N$. 

The above represents a collection of linear equations. For example, the set of equations
\begin{align}
x+2y+4z=\;&6\\
3x+2y+9z=\;&7\\
6x+4y-2z=\;&8
\end{align}
can be written in a matrix formulation as
\begin{align}\label{Axb}
\left(\begin{array}{ccc}
1 & 2 & 4\\
3 & 2 & 9\\
6 & 4 & -2
\end{array}\right)\left(\begin{array}{c}
x\\
y\\
z
\end{array}\right)=\left(\begin{array}{c}
6\\
7\\
8
\end{array}\right)
\end{align}
where we want to solve for $x$, $y$, and $z$ that satisfy all three equations simultaneously ({\it i.e.}, finding the intersection of three different lines).

Quantum physics is one area where linear algebra problems are necessary to solve in order to discover the fundamental properties of chemistry systems, solve particle physics problems, and find solutions for modelling modern technologies.

In any quantum physics problem, there is a basis set that is chosen, $\phi_i$, for $i\in\{1,\ldots,N\}$ which indexes $N$ basis functions. Beyond just a basis set, we also need the Hamiltonian function, $\hat H$, which tells us how the system behaves. Even though we can write down the Hamiltonian for a system, solving it requires advanced techniques, which is what we are preoccupied with in this paper. Often, the fundamental elements of these solvers are not covered in standard physics texts, but in order to make new advances, it is necessary to understand the fundamentals of linear algebra for quantum problems.

This paper is designed for readers who have completed a basic introduction to quantum mechanics but are curious about how numerical methods can be used to solve the underlying linear algebra problems. The optimization of linear algebra routines has been a longstanding research interest in computer science. Many decades of effort have gone into optimizing the code and also the interface with hardware. The time needed to improve upon these algorithms is lengthy, and one of our goals is to emphasize the point that the use of these efficient libraries is often better than trying to code these basic operations oneself.  However, understanding of these methods is frequently vital to understanding how to make the most efficient algorithms and make new methods to solve new problems.

\subsection{The Problem to Solve in Quantum Mechanics}

In quantum physics, the time-independent Schr\"odinger equation is
\begin{equation}\label{schrodinger}
\hat H | \mathbf{\psi} \rangle = E | \mathbf{\psi} \rangle
\end{equation}
where $\hat H$ is a Hamiltonian operator matrix, $E$ is a scalar energy eigenvalue, and $| \mathbf{\psi} \rangle$ is an energy eigenstate wavefunction (eigenvector). 

Throughout the text, we use the form for vectors that is used in quantum mechanics. A ket vector $|\psi\rangle$ will correspond to a column vector. Meanwhile, a dual vector (belonging to the dual vector space), $\langle\psi|$, will be a row vector satisfying the relation  $\langle\psi| = |\psi\rangle^{\dagger}$.

Notice that Eq.~\eqref{schrodinger} is in the same form as Eq.~\eqref{genlineq} with $\hat A=\hat H$, $\mathbf{x} = | \mathbf{\psi} \rangle$, and $\mathbf{b} = E | \mathbf{\psi} \rangle$. Though in this form,  Eq.~\eqref{schrodinger} cannot be solved in the same way as Eq.~\eqref{genlineq} since we do not know $| \mathbf{\psi} \rangle$ on either side. Thus, Eq.~\eqref{schrodinger} must be rearranged as
\begin{equation}\label{nullspace}
    (\hat H - E \mathbb{I}) | \mathbf{\psi} \rangle = \mathbf{0}
\end{equation}
where we then find what are called the nullspace vectors $| \mathbf{\psi} \rangle$ of $\hat H - E \mathbb{I}$. This is the intuitive link between these equations. Though, in practice, one would not try to solve Eq.~\eqref{schrodinger} by finding the set of nullspace vectors for Eq.~\eqref{nullspace} one by one. Instead, more elegant solution strategies from methods in linear algebra may be applied.

We note that the generalized eigenvalue problem for systems with non-orthogonal basis functions would appear as
\begin{equation}
\hat H | \mathbf{\psi} \rangle = E \hat S | \mathbf{\psi} \rangle
\end{equation}
for some overlap matrix $\hat S$ containing the inner products of all basis functions used, $S_{ij}=\langle\phi_i|\phi_j\rangle$, where this expression would be an identity matrix if all basis states were orthogonal. We will focus on systems with orthogonal bases, although linear algebra problems in quantum chemistry applications are frequently constructed where the basis functions are not orthogonal. For the interested reader, we refer to Ref.~\onlinecite{anderson1999lapack} for a full account of how to use and program algorithms for systems with non-orthogonal eigenbases. Hence, with this choice, $\hat S=\mathbb{I}$ here.

Solutions to Eq.~\eqref{schrodinger} do not completely characterize a quantum system as this is just the time-independent Schr\"odinger equation. The time-dependent solutions are given by unitary time evolution of the stationary states obtained from Eq.~\eqref{schrodinger} as

\begin{equation}\label{time-evo}
    | \Psi(t) \rangle = \hat U(t)|\psi(0)\rangle= e^{-i \hat H t / \hbar} | \psi(0) \rangle.
\end{equation}

The exponentiation of $-i \hat H t / \hbar$ can be expanded as a summation to give the unitary matrix
\begin{equation}
    \hat U(t) = e^{-i \hat H t / \hbar} = \sum_{k=0}^{\infty} \frac{1}{k!} \left(\frac{-i \hat H t}{\hbar}\right)^k
\end{equation}
which is accomplished by a series expansion of the exponential \cite{Berry}.

Performing time evolution in the basis of the Hamiltonian, such that $\hat H$ is a diagonal matrix with energies $E_k$ populating its diagonal, makes the process trivial as $\hat U(t)$ becomes
\begin{equation}
    \hat U(t) =
    \left(\begin{array}{cccc}
        e^{-i \hat E_1 t / \hbar} & 0  & \cdots & 0\\
        0 & e^{-i \hat E_2 t / \hbar } & \cdots & 0\\
        \vdots & \vdots & \ddots & \vdots\\
        0 & 0& \cdots &  e^{-i \hat E_n t / \hbar}\\
    \end{array}\right)\\
\end{equation}
and then application in Eq.~\eqref{time-evo} simply becomes elementwise multiplication.

\subsection{Linear Algebra Framework for Quantum Problems}

There are a few key concepts which require a firm grasp to build any quantum problem.

\paragraph{Quantum state.--}Any entity which can be described by quantum mechanics has a discrete set of states which it may occupy. When considering multiple entities, combinatorics are required to count the total number of possible states. The simplest example is a set of spin-1/2 particles (say electrons) which may be in one of two states with respect to a given axis: up or down. Let us disregard any other information associated with these particles and just focus on this singular quantity. For $n$ sites, we may then have $2^n$ possible states $| \uparrow \uparrow \ldots \uparrow \uparrow \rangle$, $| \uparrow \uparrow \ldots \uparrow \downarrow \rangle$, $| \uparrow \uparrow \ldots \downarrow \uparrow \rangle$, $| \uparrow \uparrow \ldots \downarrow \downarrow \rangle$, and so on. The exponential relationship based on $n$ implies that the wavefunction and the space that the problem is formulated in will grow very rapidly with the number of sites!

Part of what makes quantum mechanics so difficult to solve is that we must pick variables for the basis functions that correspond to good quantum numbers. We cannot simply include all quantum numbers in the basis state. So, in the above example we only concentrated on the $\hat z$ direction for the quantized spin. All variables in a basis function must correspond to a complete set of commuting observables \cite{townsend2000modern}. This construction allows us to obey the uncertainty principle and to keep basis functions constant in time.

\paragraph{Basis vector.--}Each of the above states is associated with a standard basis vector. If we start with $n=1$ electron, then the standard basis vectors are
\begin{equation}
| \uparrow \rangle =(1,0)^T=\left(\begin{array}{c}
1\\
0
\end{array}\right)
\end{equation}
and
\begin{equation}
| \downarrow \rangle = (0,1)^T=\left(\begin{array}{c}
0\\
1
\end{array}\right).
\end{equation} 
Then for $n=2$ electrons we might have 
\begin{equation}
| \uparrow \uparrow \rangle=|\uparrow\rangle\otimes|\uparrow\rangle = (1, 0, 0, 0)^T=\left(\begin{array}{c}
1\\
0\\
0\\
0
\end{array}\right)
\end{equation}
where $\otimes$ is a Kronecker product and $T$ is the transpose operator. The Kronecker product in this case multiplies the vector on the right onto the vector on the left's elements as
\begin{align}
A\otimes B=\left(\begin{array}{c}
\alpha_1\\
\alpha_2
\end{array}\right)\otimes\left(\begin{array}{c}
\beta_1\\
\beta_2
\end{array}\right)&=\left(\begin{array}{c}
\alpha_1\left(\begin{array}{c}
\beta_1\\
\beta_2
\end{array}\right)\\
\alpha_2\left(\begin{array}{c}
\beta_1\\
\beta_2
\end{array}\right)
\end{array}\right)\\
&=\left(\begin{array}{c}
\alpha_1\beta_1\\
\alpha_1\beta_2\\
\alpha_2\beta_1\\
\alpha_2\beta_2
\end{array}\right)
\end{align}
which effectively joins together two different Hilbert spaces. In the case of spin systems, the states of one spin with another. We can think about this more broadly as enumerating all possible states for two sites.

Similarly, $| \uparrow \downarrow \rangle = (0, 1, 0, 0)^T$, $| \downarrow \uparrow \rangle = (0, 0, 1, 0)^T$, and $| \downarrow \downarrow \rangle = (0, 0, 0, 1)^T$. Notice that while each ket contains $n$ arrows, the associated basis vector will be of length $N=2^n$ such that all states are enumerated.

Beyond spin-half systems, we can have spin systems with $d$ states on each site. In this case, we would have $N=d^n$ states for $n$ sites. In quantum chemistry problems we would choose an orbital basis $\phi_i$ that corresponds to non-orthogonal functions. 

There is no universal basis set and the choice depends on the problem of interest. Any complete, orthonormal, set of eigenvectors may be used. Any $N \times N$ operator representing some physical observable (perhaps a Hamiltonian with an observable energy) is always set in a basis with $N$ orthogonal eigenvectors \cite{Boyce}. Thus, the set of normalized eigenvectors of any Hermitian operator may be taken to form the basis. 

For some systems, multiple quantities must be considered. For example, to enumerate all the possible states of a hydrogen atom in its rest frame one must account for the distance between the electron and nucleus, the orbital angular momentum, the orbital orientation encapsulated by the magnetic quantum number, and the intrinsic spin orientation \cite{townsend2000modern}.

\paragraph{Probabilities and normalization.--}Consider being in the basis of some operator $\hat A$ with the state vector $|\mathbf{\psi} \rangle$ in some superposition of basis states. For the simple case of a single spin-1/2 particle with the basis oriented along some spin axis, we would have $|\mathbf{\psi} \rangle = (\alpha,\beta)^T$ such that $\alpha,\beta \in \mathbb{C}$ and $|\alpha|^2 + |\beta|^2 = 1$. The particle then has some probability of being in either the $| \uparrow \rangle$ state or the $|  \downarrow \rangle$ state. This non-deterministic nature lends itself to quantum computing, where unitary operators (or gates) are applied to quantum state vectors, splitting the probabilities across basis states as desired. 

\paragraph{Wavefunction.--}One can write a wavefunction into a spectral decomposition of basis states of the form
\begin{equation}
|\psi\rangle=\sum_{i=1}^N\alpha_i|\phi_i\rangle
\end{equation}
for some coefficients $\alpha_i$ corresponding to the probability amplitudes. Quantum states are those that possess a non-zero amount of entanglement entropy \cite{bakerCJP21}. So, the state cannot be a simple Kronecker product of basis states as this is not entangled. Any time that two of the $\alpha_i$ coefficients are non-zero, we would call this a superposition of basis states. From the wavefunction, the observed quantities of a quantum system can be obtained.

\paragraph{Operator.--} When we say operator, we can, for the purposes of this paper, just think of a matrix. The form of the operators that we use come directly from quantum physics, including the most prominent operator: the Hamiltonian. In general, an operator acts on the wavefunctions.

The focus of this paper is on computational quantum problems, which often boil down to correctly describing a system by a Hermitian Hamiltonian then finding the allowed energy eigenstates. Once systems become too large other strategies are required, but methods from computational linear algebra still lay at the core of any technique that is based on the construction here.

\subsection{Computational Complexity Measured by Big-$O$ Notation}

There are a variety of strategies that can be used to solve Eq.~\eqref{genlineq}. A unifying tool to analyze the performance of these linear algebra algorithms is big-$O$ notation, which is formally used to denote an asymptotic upper bound on the worst case time or space complexity of an algorithm~\cite{Goodrich}.\footnote{Big-$O$ notation is not the only tool for analyzing algorithmic complexity. There are also little-$o$ (strict upper bound), big-$\Theta$ (asymptotic tight bound), big-$\Omega$ (asymptotic lower bound), and little-$\omega$ (strict lower bound)~\cite{Goodrich}.} The quantity inside the $O$ symbol corresponds to the worst case number of operations (i.e.~running time) or amount of memory required to run an algorithm with respect to input size where any constant prefactors are dropped (e.g.~$O(cn^2)=O(n^2)$).\footnote{It is not always the case that big-$O$ notation is used to denote the worst case. In the case of randomized algorithms, some authors use big-$O$ notation to denote an upper bound on the expected running time of the algorithm, whereas the true worst case running time might be much worse. It is often stated explicitly when this is done.} Often the number of operations for a given step are equated. So, for example, if three loops have a number of iterations of $M$, $N$, and $P$, with $N \geq M$, $N \geq P$, then the resulting complexity is expressed in big-$O$ notation as $O(N^3)$ instead of $O(NMP)$ to more conveniently represent the worst-case evaluation of the algorithm. This helps with comparing algorithms to each other.

Big-$O$ notation can be applied as a comparator to understand how multiple algorithms should behave with respect to one another. However, one must be aware that when comparing two algorithms big-$O$ notation can leave out important information. One can imagine the computational benefits which could be incurred if an operation of $O(N^{\alpha})$ could be replaced by one of $O(N^{\beta})$ where $\alpha > \beta$. However, it is important to remember that constant factors and sub-leading terms are hidden by big-$O$ notation such that an algorithm of time-complexity $O(N^{\beta})$ may only run in fewer operations for impractically large $N$. Thus, the cross-over point between two algorithms might be only for very large $N$. It is always wise to be cautious of many theoretical algorithms that boast low worst-case time-complexities but have no actual use cases due to memory limitations, large pre-factors hidden in the complexity scaling, or difficulty of implementation.

\subsection{Gaussian Elimination}

Gaussian Elimination is a fundamental technique for the solution of Eq.~\eqref{genlineq}. This method transforms an augmented matrix (a coefficient matrix along with its constant vector) into row-echelon form in which the matrix is transformed to triangular form while the constant vector is transformed in tandem by relevant operations, from which back-substitution can subsequently be used to find a solution vector for Eq.~\eqref{genlineq}. It is widely used within more advanced algorithms, and so we discuss it here before motivating the problem from a different angle that is more common in linear algebra solvers. 

Gaussian elimination involves performing three elementary row operations (row swapping, non-zero row scaling, and addition of one row to another) algorithmically to obtain a triangular matrix \cite{Chapra-Canale}. The entire algorithm runs in $\frac{2}{3} N^3$ ($+O(N^2)$ for the backwards step) floating point operations (flops; a flop represents the unit arithmetic operation in a computer) \cite{Trefethen-Bau}.\footnote{Depending on matrix structure this cost may be reduced, e.g. for Hermitian positive definite matrices various matrix decompositions can be found in half the number of flops.}

There is a nuance when performing Gaussian elimination. Expressing Eq.~\eqref{Axb} in augmented matrix form, we can write
\begin{equation}\label{augmentedmatrix}
\hat A \mathbf{x} = \mathbf{b}\Rightarrow
\left(\begin{array}{ccc|c}
1 & 2 & 4 & 6\\
3 & 2 & 9 & 7\\
6 & 4 & -2 & 8
\end{array}\right)
= [ \hat{A} | b ]
\end{equation}
which in this form when one row times a number is equal to another, the rows can be said to be parallel. The notion of a matrix rank\footnote{Not to be confused with the rank of a tensor, which is the number of indices on the tensor.}, $\mathrm{rank(\hat A)}$, ({\it i.e.}, equal to the dimension of the vector space--in Eq.~\eqref{augmentedmatrix} it is 3) can be formalized. When the rank of the augmented matrix $[ \hat{A} | b ] $ is equal to the number of variables $N$ (i.e.~rows), that is, $N=\mathrm{rank}([ \hat{A} | b ])$, then a solution exists and is unique (i.e.~the number of unknowns equals the matrix rank). This is a core statement behind the Rouché-Capelli theorem~\cite{Shafarevich}. If instead $N > \mathrm{rank}([ \hat{A} | b ])$, then there are an infinite number of solutions and the system is said to be underdetermined. A system is said to be overdetermined when $N < \mathrm{rank}([ \hat{A} | b ])$, as there are more equations than unknowns, and often has no solution, but can have a solution if all the equations are consistent. However, for certain computational methods, it is possible for an overdetermined consistent system to become inconsistent due to the accumulation of errors via the solution process. Methods have been developed to overcome this issue and improve the accuracy for solutions to consistent overdetermined systems~\cite{Cho}.

The naive Gaussian elimination algorithm works on the $k^{th}$ iteration as follows: take augmented matrix element $a_{kk}$ (called the pivot) and for each row $j$, $j>k$, subtract row $k$ times $x_{jk} / x_{kk}$ from row $j$~\cite{Trefethen-Bau}. The result is that all rows below row $k$ will be zero-valued in column $k$.

Naive Gaussian elimination is prone to numerical errors, especially if any pivot $x_{kk}$ is near-zero valued. Accuracy and stability can be improved through techniques such as total or partial pivoting~\cite{Trefethen-Bau}. In total pivoting (also called complete pivoting), at each step the entire remaining submatrix is searched for the element of the greatest absolute value, then columns and rows are swapped to place that element in location $(k,k)$. Selecting the pivot in this way greatly reduces numerical instability, but is quite costly requiring $O(N^3)$ additional operations throughout the entire algorithm. Partial pivoting involves searching for the element of the greatest absolute value in just the $k$th column and swapping rows such that that element is placed in position $(k,k)$. Partial pivoting also reduces numerical instability and costs just $O(N^2)$ additional operations. For these reasons, partial pivoting is often applied in practice as it runs much faster when compared against total pivoting and the losses in stability are often only marginal~\cite{Trefethen-Bau}.

It should be noted that division by a near-zero floating point number is still possible with partial pivoting, just less likely to occur. For this reason, some applications may favor the use of total pivoting.

\section{Matrix Multiplication}

Let's take a moment to study and understand how one of the most ubiquitous linear algebra techniques (matrix multiplication) is implemented on a computer. The purpose of this is to lay the groundwork for how we will talk about other methods in linear algebra.

Given matrices $\hat A$ and $\hat B$, we may express them as
\begin{equation}
\hat A=\left(\begin{array}{cccc}
a_{11} & a_{12} & a_{13} & \cdots\\
a_{21} & a_{22} & a_{23} & \cdots\\
a_{31} & a_{32} & a_{33} & \ddots\\
\vdots & \vdots & \ddots & \ddots
\end{array}\right)=\sum_{ij}a_{ij}|i\rangle\langle j|
\end{equation}
and
\begin{equation}
\hat B=\left(\begin{array}{cccc}
b_{11} & b_{12} & b_{13} & \cdots\\
b_{21} & b_{22} & b_{23} & \cdots\\
b_{31} & b_{32} & b_{33} & \ddots\\
\vdots & \vdots & \ddots & \ddots
\end{array}\right)=\sum_{ij}b_{ij}|i\rangle\langle j|
\end{equation}
where $\hat A$ is a matrix with dimensions $M\times P$ and $\hat B$ is a matrix with dimensions $P\times N$. 

Matrix multiplication is then defined as
\begin{equation}\label{matmult}
\hat C = \hat A \hat B = \sum_{i=1}^M \sum_{j=1}^N \sum_{k=1}^P a_{ik} b_{kj} |i\rangle\langle j|
\end{equation}
where the resulting matrix $\hat C$ is of size $M \times N$ with the inner dimension $P$ being contracted out. Another way to view this is that each element of $\hat C$ is given by
\begin{equation}
c_{ij}=\sum_{k=1}^P a_{ik}b_{kj}.
\end{equation}

Visually, we can see that the rows of $\hat A$ and the columns of $\hat B$ are multiplied together as
\begin{align}
\hat A \hat B &=
\begin{bmatrix}
    \tikz[baseline=(X.base)]\node[draw=blue,fill=white,circle,inner sep=1pt] (X) {$a_{11}$}; & \tikz[baseline=(X.base)]\node[draw=orange,fill=white,circle,inner sep=1pt] (X) {$a_{12}$}; & \tikz[baseline=(X.base)]\node[draw=green,fill=white,circle,inner sep=1pt] (X) {$a_{13}$}; \\
    a_{21} & a_{22} & a_{23} \\
    a_{31} & a_{32} & a_{33}
\end{bmatrix}
\begin{bmatrix}
    \tikz[baseline=(X.base)]\node[draw=blue,fill=white,circle,inner sep=1pt] (X) {$b_{11}$}; & b_{12} & b_{13}\\
    \tikz[baseline=(X.base)]\node[draw=orange,fill=white,circle,inner sep=1pt] (X) {$b_{21}$}; & b_{22} & b_{23} \\
    \tikz[baseline=(X.base)]\node[draw=green,fill=white,circle,inner sep=1pt] (X) {$b_{31}$}; & b_{32} & b_{33}
\end{bmatrix}
\\
&=
\begin{bsmallmatrix}
   \tikz[baseline=(X.base)]\node[draw=blue,fill=white,circle,inner sep=1pt,scale=0.7] (X) {$a_{11}b_{11}$}; + \tikz[baseline=(X.base)]\node[draw=orange,fill=white,circle,inner sep=1pt,scale=0.7] (X) {$a_{12}b_{21}$}; +\tikz[baseline=(X.base)]\node[draw=green,fill=white,circle,inner sep=1pt,scale=0.7] (X) {$ a_{13}b_{31}$}; & a_{11}b_{12} + a_{12}b_{22} + a_{13}b_{32} & \ldots\\
    a_{21}b_{11} + a_{22}b_{21} + a_{23}b_{31} & a_{21}b_{12} + a_{22}b_{22} + a_{23}b_{32} & \ldots\\
    a_{31}b_{11} + a_{32}b_{21} + a_{33}b_{31} & a_{31}b_{12} + a_{32}b_{22} + a_{33}b_{32} & \ldots
\end{bsmallmatrix}.\nonumber
\end{align}

\subsection{Implementation and Scaling of Matrix Multiplication}

The computational algorithm to multiply two matrices is shown in Code Block~\ref{code:mult}. Three loops are required to execute the algorithm, yielding a worst-case running time of $O(N^3)$ where $N$ is the largest matrix-dimension involved.

\begin{lstlisting}[language=Julia, caption={Matrix multiplication}, label={code:mult}]{code_examples/matmult.jl}{julia}
function matmult(A, B)
  N = size(A,1)
  M = size(B,2)
  P = size(A,2)
  C = Array{Float64,2}(undef,M,N)
  for i = 1:N
    for j = 1:M
      out = 0
      for k = 1:P
        out += A[i,k]*B[k,j]
      end
      C[i,j] = out
    end
  end
  return C
end
\end{lstlisting}

To give another example of scaling, the addition of two matrices would require us to sum over indices $i$ and $j$ as
\begin{equation}
\hat A + \hat B = \sum_{i=1}^M \sum_{j=1}^N (a_{ij}+b_{ij}) |i\rangle\langle j|
\end{equation}
requiring $O(MN)$ (and where $O(N^2)$ would be the normally reported case if $M=N$) flops since there are two for-loops.

In comparing these time-complexities, one would say that the cost of adding two matrices is smaller than their multiplication by a factor of $N$.

\begin{figure*}
    
    \begin{subfigure}{0.45\columnwidth}
        \includegraphics[width=\linewidth]{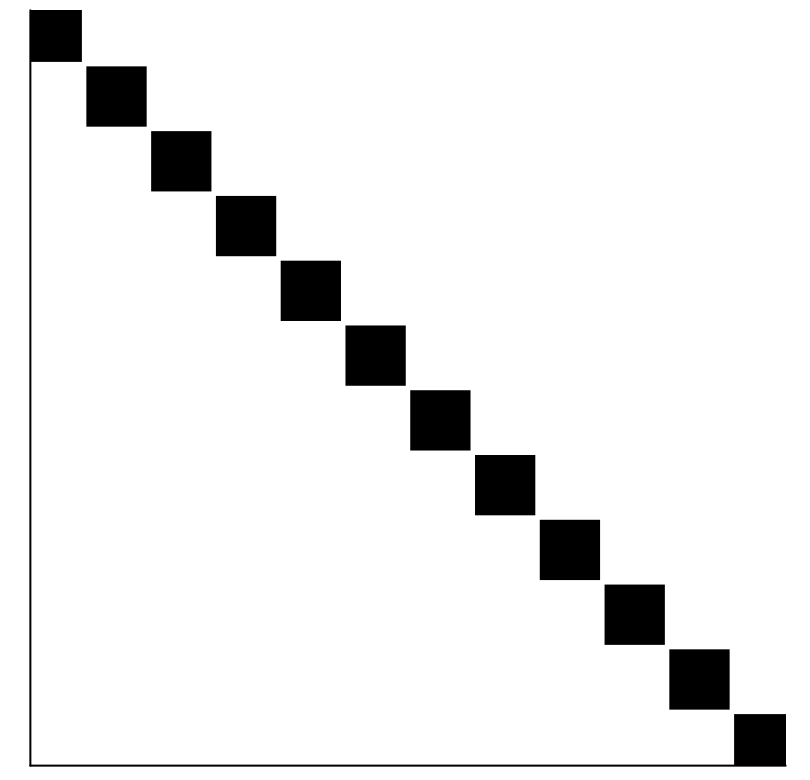}
        \caption{}
    \end{subfigure}
    \begin{subfigure}{0.45\columnwidth}
        \includegraphics[width=\linewidth]{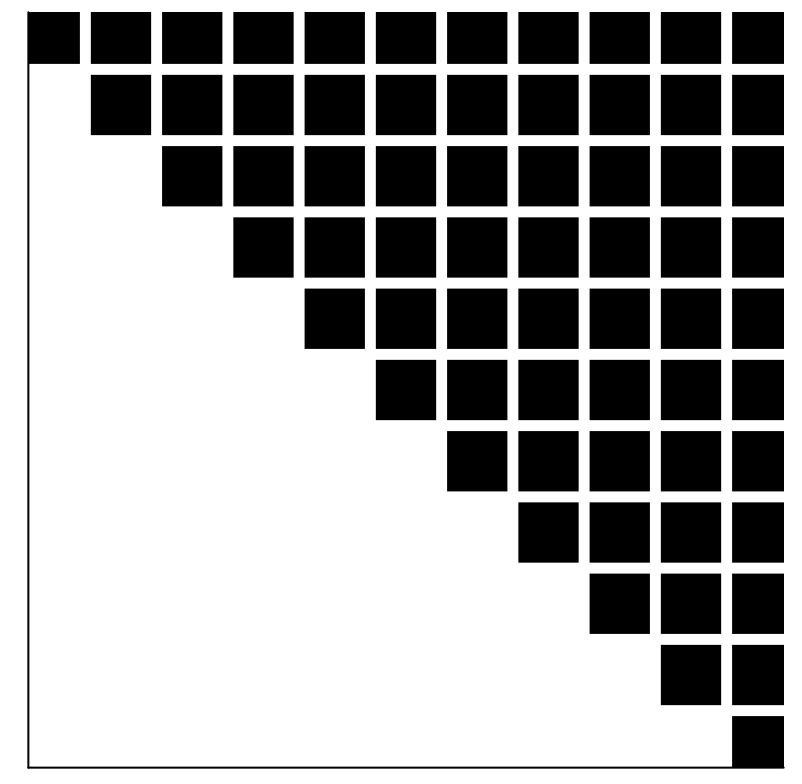}
        \caption{}
    \end{subfigure}
    \begin{subfigure}{0.45\columnwidth}
        \includegraphics[width=\linewidth]{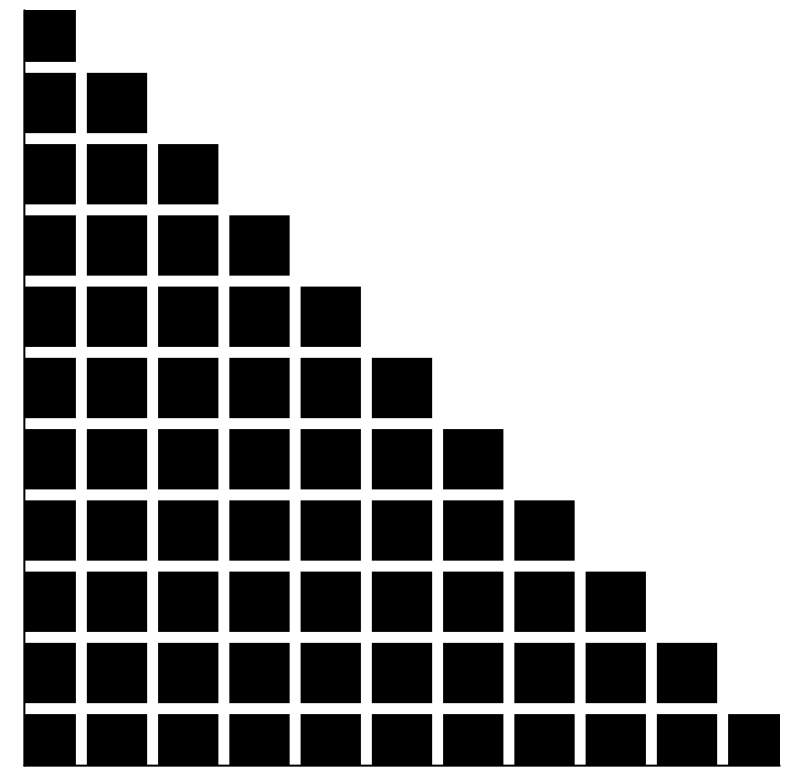}
        \caption{}
    \end{subfigure}
    \begin{subfigure}{0.50\columnwidth}
        \includegraphics[width=\linewidth]{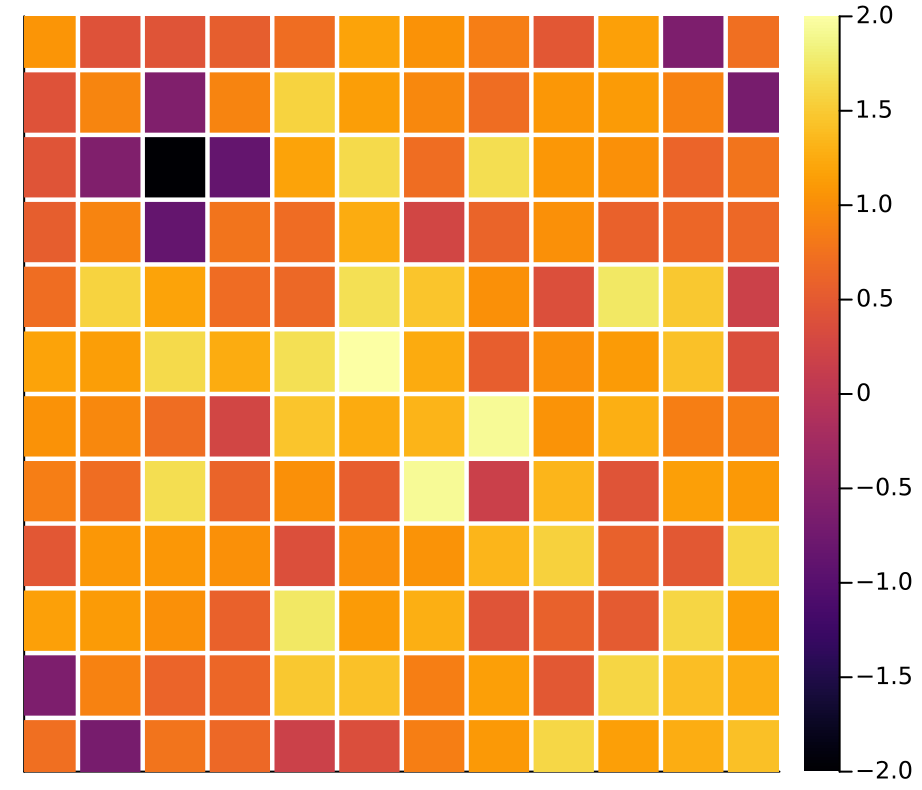}
        \caption{}
    \end{subfigure}
    \begin{subfigure}{0.45\columnwidth}
        \includegraphics[width=\linewidth]{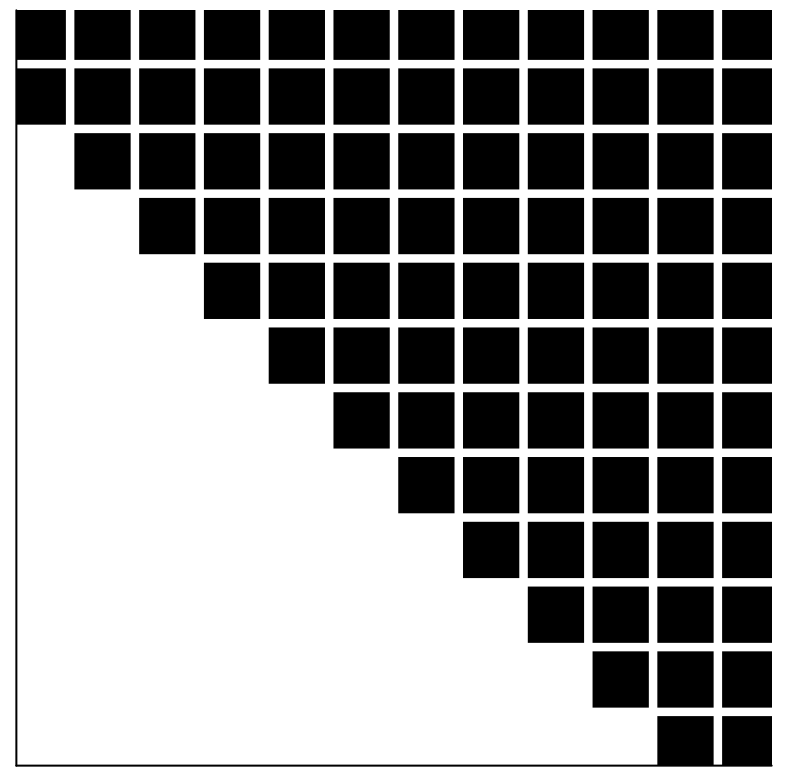}
        \caption{}
    \end{subfigure}
    \begin{subfigure}{0.48\columnwidth}
        \includegraphics[width=\linewidth]{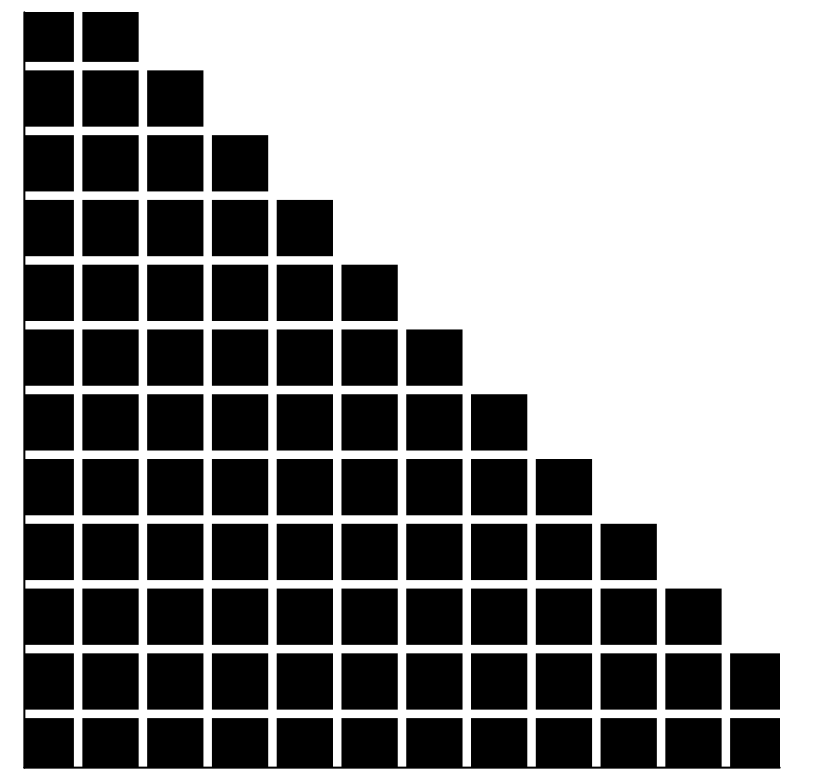}
        \caption{}
    \end{subfigure}
    \caption{Visual representations of matrices: (a) an identity matrix, (b) an upper triangular matrix, (c) a lower triangular matrix, (d) a symmetric matrix, (e) an upper Hessenberg matrix, and (f) a lower Hessenberg matrix.}
    \label{fig:matrices}
\end{figure*}

\subsection{Why Basic Matrix Multiplication is Often Not as Fast as Linear Algebra Libraries}

Implementations of even these elementary operations are not as simple as just writing code. The code used for matrix multiplication in Code Block~\ref{code:mult} is not as fast as an internal implementation, even though the algorithmic steps are the same. Methodologies for implementing matrix multiplication can be found in Ref.~\onlinecite{Kelefouras}.

To fully implement a fast matrix multiplication, one must take into account how the computer is constructed and how it operates. To make any algorithm maximally efficient, one has to consider the computer architecture, hardware, instructions sets, etc. Implementations improve performance beyond what is shown in Code Block~\ref{code:mult} through various techniques including efficient memory cache usage \cite{kowarschik2003overview}, architecture-specific tuning (potentially even at an assembly language level), as well as memory prefetching and loop unrolling to improve memory access patterns and ensure an efficient transfer of information to the CPUs \cite{whaley2005minimizing, vanLoan}. Adding memory pre-fetching instructions to the CPU ensures an efficient transfer of information from the hardware cache to the CPUs performing the computation.

The standard for basic linear algebra routines, including matrix-matrix multiplication, are known as the Basic Linear Algebra Subprograms (BLAS) \cite{Geijn}. Various highly optimized BLAS libraries exist such as the Intel Math Kernel Library (MKL), AMD Optimizing CPU Libraries (AOCL), IBM's Engineering and Scientific Subroutine Library (ESSL), and many more. We point the interested reader to Ref.~\onlinecite{vanLoan} for details on how to write code as efficiently as what you expect from a library like BLAS or linear algebra package (LAPACK).

\section{Matrix Properties}

The matrix multiplication discussion above is only for dense matrices where all entries are non-zero. Some other types of matrices allow for reduced computational scaling for matrix multiplication and also can be useful for decomposing matrices, which we will consider afterward.

Matrices can be categorized into different types based on their entries. In this section, we go over what kinds of matrices become important in the algorithms that follow. Visualizations of some of these matrix types can be seen in Fig.~\ref{fig:matrices}.

A matrix can be either dense or sparse, which are not technically rigorous definitions, but are useful in practice. A dense matrix has mostly non-zero entries, while a sparse matrix has mostly zero-valued entries. In this work, a matrix is sparse if it has under 10\% to 15\% non-zero entries. Other works define a sparse matrix as one where the number of non-zero entries is linear in the matrix dimension, \textit{i.e.} $O(N)$ for an $N \times N$ matrix \cite{Ascher}.

A common theme in computational linear algebra methods is utilizing matrix structure\footnote{By matrix structure, we mean sparse matrix types, Toeplitz, Hankel, and other structures. Using these specific types can greatly decrease the complexities of many operations. For more information on specific structures and how reductions in operations for various tasks can be performed, we refer to Refs.~\onlinecite{
chan1996conjugate,ng2004iterative,garoni2017generalized,garoni2018generalized}.} to obtain greater efficiency. Depending on the algorithm, it may be beneficial to convert a matrix into a selected form before giving it as input. Many matrix structures exist beyond those mentioned here.

\subsection{Identity}

The identity matrix (Fig.~\ref{fig:matrices}a) is a square matrix with ones along its diagonal and zeros elsewhere. Matrix multiplication by the identity is analogous to scalar multiplication by 1. According to category theory, the identity matrix of size $N \times N$ is the identity morphism mapping any matrix with $N$ rows to itself. It is often denoted by $\mathbb{I}$.

Any matrix to the power of zero results in an identity matrix. The determinant of an identity matrix is equal to 1, implying that there exists an inverse matrix of the identity matrix, which is just the identity matrix.

\subsection{Triangular}

There are two classes of triangular matrices: upper triangular matrices (Fig.~\ref{fig:matrices}b) which are a form of square matrix with non-zero values only on and above the main diagonal of the matrix; and lower triangular matrices (Fig.~\ref{fig:matrices}c) which similarly only have non-zero entries on or below the main diagonal. These forms can be seen in Fig.~\ref{fig:matrices}, but it is not necessary that all entries in the dark region be non-zero.

Triangular matrices have the following useful properties \cite{Golub-vanLoan}: computing the inverse of an upper (lower) triangular matrix yields an upper (lower) triangular matrix, if one exists; the sum or product of any two upper (lower) triangular matrices is also upper (lower) triangular; and the eigenvalues of a triangular matrix are just its diagonal entries.

Additionally, Eq.~\eqref{genlineq} can be solved in $O(N^2)$ time via a method known as back substitution for upper triangular matrices or forward substitution for lower triangular matrices \cite{Golub-vanLoan}.

\subsection{Symmetric}

Symmetric matrices (Fig.~\ref{fig:matrices}d) have entries that are symmetric across the main diagonal such that $a_{ij}=a_{ji}$ for all indices. If two symmetric matrices commute ($[\hat A,\hat B]=0$) such that $\hat A \hat B = \hat B \hat A$, then their product is a symmetric matrix \cite{Hogben}.

\subsection{Hermitian}

A matrix is Hermitian if its conjugate transpose is equal to the original matrix ($\hat A^\dagger=(\hat A^*)^T=\hat A$). Hermitian matrices may have real or complex values, but the diagonal entries must all be real-valued. The real part of a Hermitian matrix is symmetric while the imaginary part is anti-symmetric (also called skew-symmetric).

Hermitian matrices can be diagonalized by unitary matrices in an eigenvalue preserving similarity transformation to create a real-valued diagonal matrix. This implies that the eigenvalues of a Hermitian matrix are real-valued. The sum of two Hermitian matrices results in a Hermitian matrix. Furthermore, the inverse of a Hermitian matrix is also Hermitian.

In general, one can write a Hermitian matrix as a symmetric real matrix plus an anti-symmetric imaginary matrix. The following is an example in the 2 by 2 case.

\begin{align}
\hat H &= 
\left(\begin{array}{cc}
a & c\\
c & b\\
\end{array}\right)
+i
\left(\begin{array}{cc}
0 & -d\\
d & 0\\
\end{array}\right)\\
\hat H^T &= 
\left(\begin{array}{cc}
a & c\\
c & b\\
\end{array}\right)
+i
\left(\begin{array}{cc}
0 & d\\
-d & 0\\
\end{array}\right)\\
\hat H^\dagger&= 
\left(\begin{array}{cc}
a & c\\
c & b\\
\end{array}\right)
+i
\left(\begin{array}{cc}
0 & -d\\
d & 0\\
\end{array}\right)\\
\hat H^\dagger &= \hat H
\end{align}

One important property of Hermitian matrices is that their eigenvalue decompositions may be expressed with unitary matrices $\hat V^{\dagger} = \hat V^{-1}$ such that $\hat H = \hat V \hat D \hat V^{\dagger}$, which is not true for any square matrix in general \cite{Garcia-Horn}.

\subsection{Unitary}

Unitary matrices are square matrices with the crucial property that $\hat U^\dagger \hat U = \mathbb{I}$, which implies $\hat U^\dagger = \hat U^{-1}$ and $\det(\hat U)=1$. We are often concerned with the real-valued rotation matrices.

\begin{equation}
    \hat U^{2 \times 2}_{R} = \left(\begin{array}{cc}
        \cos\theta & \sin\theta  \\
        -\sin\theta  & \cos\theta 
    \end{array}\right)
\end{equation}
Although more generally, we have $2 \times 2$ rotation matrices in $\mathrm{SU}(2)$ as

\begin{equation}
    \mathrm{SU}(2) = \left(\begin{array}{cc}
        \alpha &  -\beta^*  \\
        \beta  &  \alpha^* 
    \end{array}\right)
\end{equation}
where $\alpha,\; \beta \in \mathbb{C}$ and $|\alpha|^2 + |\beta|^2 = 1$ \cite{Hall}. Square unitary matrices of any dimensionality exist.

Note that the direct sum of any two matrices defined as
\begin{equation}
\hat A\oplus\hat B=\left(\begin{array}{cc}
\hat A & 0\\
0 & \hat B
\end{array}\right).
\end{equation}

Some important aspects of unitary transformations are: the multiplication of two unitary matrices yields a unitary matrix; 
the kronecker product of any unitary matrix with an identity matrix also yields a unitary matrix; the direct sum of any two unitary matrices is also unitary; the eigenvalues of a matrix stay the same after a unitary similarity transformation is applied; and applying a similarity transformation with unitaries to any Hermitian matrix yields a Hermitian matrix.

\subsection{Hessenberg Matrices}

Hessenberg matrices are square matrices with all zero-valued entries below the first sub-diagonal (upper Hessenberg,  Fig.~\ref{fig:matrices}e) or above the first super-diagonal (lower Hessenberg,  Fig.~\ref{fig:matrices}f).

Hessenberg matrices are favorable in some computations when triangular matrices cannot be obtained. In particular, we will see this is the case with the QR algorithm for finding the eigenvalues of a matrix. It generally costs fewer operations to convert any matrix to Hessenberg form then apply a variation of the QR algorithm than it is to simply run the QR algorithm on the given matrix.

Before covering the case of the Hessenberg QR algorithm, one can consider the simple example of matrix multiplication between two upper Hessenberg matrices as an analogy. Due to the relatively large number of zero-valued elements, fewer operations are required when compared against traditional matrix multiplication.

\begin{lstlisting}[firstline=3, lastline=23, tabsize=2, language=Julia, caption={Matrix multiplication for Hessenberg matrices}, label={code:hessenberg}]{julia}
using LinearAlgebra

function upper_Hessenberg_mult(A, B)
  x = size(A,1)
	y = size(B,2)
	z = size(A,2)
	C = zeros(x,y)
	for i in 1:x
		j_start = i<=2 ? 1 : i-2
		for j in j_start:y
			out = 0
			k_start = i==1 ? 1 : i-1
			for k in k_start:j
				out += A[i,k]*B[k,j]
			end
			if j < z
				out += A[i,j+1]*B[j+1,j]
			end
			C[i,j] = out
		end
	end
	return C
end

A = zeros(100, 100)
B = zeros(100, 100)

for i in 1:size(A)[1]
	for j in 1:size(A)[2]
		if i-2 < j
			A[i, j] = rand()[1]
		end
	end
end

for i in 1:size(B)[1]
	for j in 1:size(B)[2]
		if i-2 < j
			B[i, j] = rand()[1]
		end
	end
end

display(norm(upper_Hessenberg_mult(A, B) - A*B))
\end{lstlisting}

One especially important note is that any Hermitian matrix transformed into a Hessenberg form via unitary similarity transformations will end up being in tridiagonal form, as we will see in Sec.~\ref{HessenbergTransform}.

\subsection{Sparse Matrices}

In this work, a sparse matrix is considered to be one with only $10\%$ to $15\%$ non-zero entries, excluding the identity matrix due to its special properties. In the context of quantum mechanical operators, a sparse matrix could indicate interactions/couplings between relatively few basis states.

Often in quantum mechanics, we deal with sparse Hamiltonians because we expect that interactions are local, and therefore the Hamiltonian does not couple all sites to all other sites. These may be exactly or numerically solvable. A sparse Hamiltonian can be used to approximate a system then have some perturbative Hamiltonian added to it, resulting in a Hamiltonian with more non-zero entries. This result may still be sparse depending on the perturbation chosen. Computational methods may be applied onto the perturbed system using solutions to the unperturbed system as a first approximation. 

\subsubsection{Hamiltonian of a Spin Matrix}

A Hamiltonian in quantum mechanics encapsulates the behavior of the system in question. There are a few general properties of any Hamiltonian in quantum mechanics. For what we consider, the Hamiltonian is Hermitian.

\begin{figure}   
    \includegraphics[width=\columnwidth]{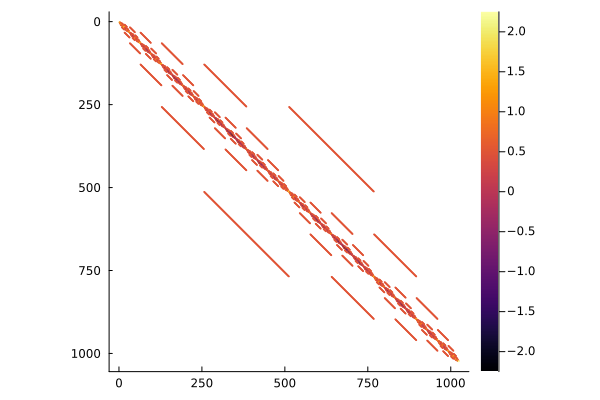}
    \caption{A spin-1/2 XXZ Heisenberg model with 10 sites and therefore $2^{10}=1024$ possible states. Each colored index indicates some coupling between states (consider applying this matrix to a quantum state vector).  This matrix is sparse and Hermitian.}
    \label{fig:heisenbergmodel}
\end{figure}

\begin{figure}
    \includegraphics[width=\columnwidth]{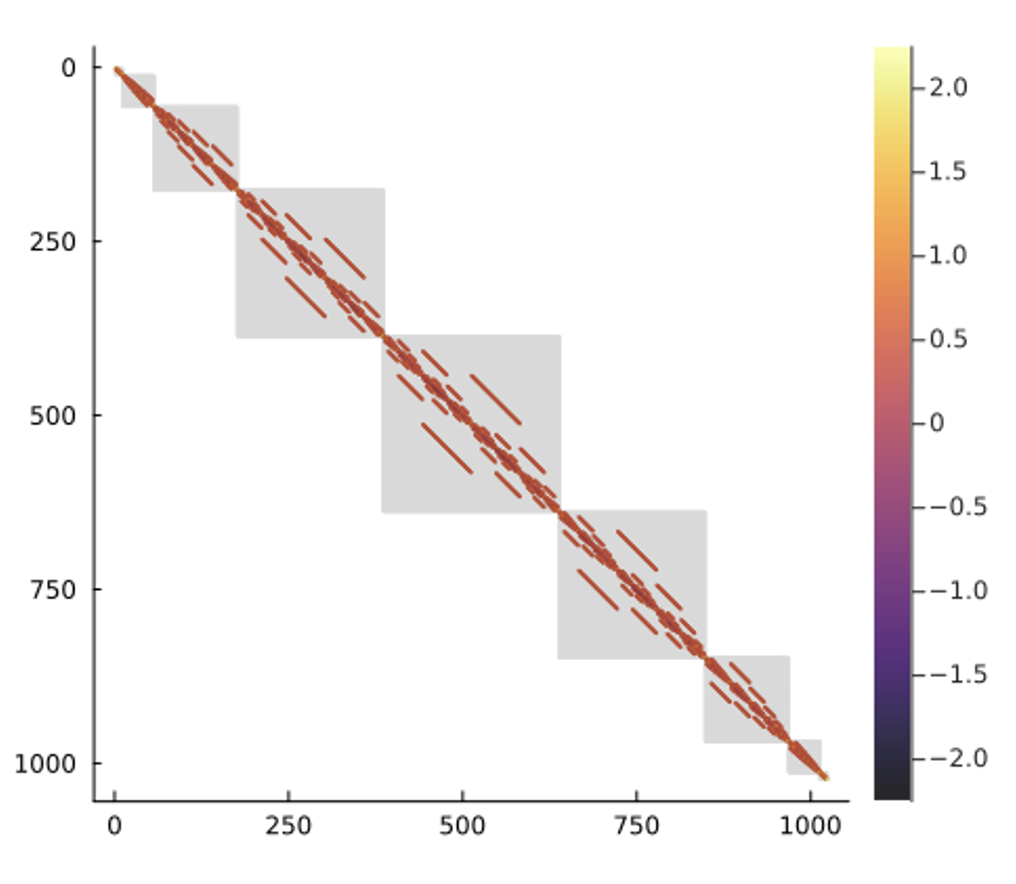}
    \caption{The Hamiltonian of a 10-site spin-half Heisenberg model such that each row and column on the Hamiltonian is associated with a specific quantum number relating to the total spin moment of the system. These quantum numbers are grouped together into blocks from +10 to -10 in integer steps. These blocks are what are displayed on the graph and show the structure of the symmetries in the problem.
    }
    \label{fig:heisenbergblocks}
\end{figure}

The Hamiltonian describing a 10-site spin-1/2 XXZ Heisenberg model in the absence of an external magnetic field with unit coupling between sites is
\begin{equation}
\hat H = \sum_i\frac12\left(\hat S^+_i \hat S^-_{i+1}+\hat S^-_{i+1}\hat S^+_i\right)+\Delta \hat S^z_i \hat S^z_{i+1}
\end{equation}
for some real number $\Delta$. The matrix elements of $\hat H$ are shown in Fig.~\ref{fig:heisenbergmodel}. 

Each of the physical indices has a set of quantum numbers that correspond to each value on the index. For the spin model it would be $\sigma=\uparrow,\downarrow$. When Kroneckered together into the full vector space, the indices sum together according to the rules of the group. In order to recover the blocks in Fig.~\ref{fig:heisenbergmodel}, we compute the sum of all quantum numbers and group common sums together. This gives us an image that goes from all spins up (+10, given in units of a half-spin) to all spins down (-10). The central block is the largest (+0) containing states with no polarization.

Notice that the blocks are independent of each other. There is no term that couples two independent blocks together. This form can be used in algorithms. Notably, each of these blocks can be handled independently allowing us to leverage this matrix substructure for faster computations.

There are other types of quantum symmetries in problems that we would like to solve. They all have the common feature that the blocks are independent and do not transfer information outside of the set of the states which they may act on.

\subsubsection{Block Matrices}

Square and rectangular matrices can be partitioned into block matrices arbitrarily \cite{Anton-Rorres}. Matrix partitioning plays a key role in some algorithms, often towards improving efficiency. Fig.~\ref{fig:heisenbergblocks} exhibits block structure. Below is an example of how some $3 \times 3$ matrix might be partitioned.

\begin{align}
    \hat H &=
    \left(\begin{array}{ccc|c}
        a & b & c & d\\
        e & f & g & h\\
        i & j & k & l\\
        \hline
        m & n & o & p\\
    \end{array}\right)\\
    \hat H &=
    \left(\begin{array}{c|c}
        A & B\\
        \hline
        C & D\\
    \end{array}\right)
\end{align}

By carefully choosing a matrix partitioning this way, one may be able to compute a solution to a system of equations much more efficiently depending on the structure. If a matrix is known in advance to have large blocks which are all zero-valued, then the total number of operations can often be reduced by skipping computations for those blocks.

Speedups can also be found through specialized blocking methods for dense matrices with non-zero blocks. For example, the Strassen algorithm for matrix multiplication and its variants partition input matrices into blocks, then apply a set of equations requiring more block-matrix additions but fewer block-matrix multiplications when compared against traditional matrix multiplication \cite{Huang}. The result is a matrix multiplication algorithm with a lower time-complexity of $O(N^{\log_2 7})$ when applied recursively, though this comes at the cost of significant constant factors and sub-leading terms \cite{Huang}.

\subsubsection{Banded Matrices}

A banded matrix is a sparse matrix with non-zero entries only near the main diagonal. A banded matrix has upper and lower bandwidths according to
\begin{align}
    a_{ij} = 0\quad \forall\; (j > i+q) \lor (i > j+p)
\end{align}
where $q$ is the upper bandwidth and $p$ is the lower bandwidth \cite{Golub-vanLoan}. Banded matrix structure can be leveraged in computational linear algebra problems for efficient computation. Figs.~\ref{fig:heisenbergmodel} and \ref{fig:heisenbergblocks} exhibit a banded structure.

One simple example of a banded matrix is a tridiagonal matrix with non-zero entries only on the main, sub, and super diagonals. This form occurs for the quantum mechanical system describing a single free particle in one dimension with plane wave solutions. The one dimensional time independent Schr\"odinger equation for a free particle is
\begin{equation}
    \frac{d^2 \psi}{dx^2} = - \frac{2 m E}{\hbar^2} \psi
\end{equation}
which can be expressed in discretized space with the second order central derivative approximation as
\begin{equation}\label{oneDSchrodinger}
    \frac{1}{\Delta x^2}\left(\begin{array}{ccccccc}
    -2 & 1 & 0 & 0 & \cdots & 0\\
    1 & -2 & 1 & 0 & \cdots & 0\\
    0 & 1 & -2 & 1 & \cdots & 0\\
    0 & 0 & 1 & -2 &\ddots & \vdots\\
    \vdots & \vdots & \vdots & \ddots & \ddots & 1\\
    0 & 0 & 0 & \cdots & 1 & -2
    \end{array}\right) | \psi \rangle
    = -\frac{2 m E}{\hbar^2} | \psi \rangle
\end{equation}
where $\Delta x$ is the grid spacing along the $x$-axis. Note that this is of the form of a Toeplitz matrix. Efficient methods exist to take advantage of this structure such as matrix vector multiplication in $O(N \log N)$ flops or fast methods for the d-dimensional Laplacian, see Refs.~\onlinecite{
chan1996conjugate,ng2004iterative,garoni2017generalized,garoni2018generalized}.

In practice, tridiagonal matrices rarely occur in quantum problems. However, quantum operators can be transformed to tridiagonal form as we will see in Sec.~\ref{QRalg} and Sec.~\ref{lanczos}.

\subsubsection{Storage Formats}

Sparsity can be leveraged such that less memory is consumed through clever storage methods. Any alternative storage method requires modifications to linear algebra programs such that the benefits may be reaped. If done well, these methods will often cost fewer operations.

The most obvious format is to store elements as a list of 3-tuples where each non-zero entry has its value, row index, and column index stored. There are more memory-efficient options, however. A useful variation on this idea comes from compressed sparse row (CSR) format.

In CSR, all elements are ordered row by row which causes storing information about row indices to be redundant. In this format, we have a list with the number of non-zero elements in each row, and a list of 2-tuples containing each non-zero element and its corresponding column \cite{Saad}. To read off elements from the matrix, one would read the first integer in the row list, then pop off that many elements from the tuple list (which all exist in the first row). The next integer in the row list would be read, and the process would be repeated until all lists are exhausted.

Many libraries support standard methods like CSR. Other sparse storage methods exist such as for banded matrices or block matrices. These can be problem specific and may require unique implementations for data types and any related linear algebra methods.

\subsection{Condition Number of a Matrix}

The stability of the matrix is characterized by what is called the condition number \cite{Trefethen-Bau}. For a given matrix norm, the condition number of a matrix is defined as
\begin{equation}
    \mathrm{cond}(\hat A) = \|\hat A\|\|\hat A^{-1}\|
\end{equation}

A perfectly conditioned matrix has condition number 1. Notice that all unitary matrices, which have unit determinant and a unitary inverse, must then have a condition number of 1.

For large condition numbers, small perturbations (from rounding, numerical errors, problem formulation approximations, etc.) in the input vector which the matrix is applied to can result in relatively large deviations in the resultant vector (ill-conditioned behavior).

We are often concerned with the 2-norm condition number which simplifies the formula to the ratio between the largest and smallest singular values of $\hat A$ \cite{Golub-vanLoan}.
\begin{equation}
    \mathrm{cond_2}(\hat A) = \frac{\sigma_{max}}{\sigma_{min}}
\end{equation}
In the case of a Hermitian matrix, the singular values are the same as the absolute values of the eigenvalues \cite{Trefethen-Bau}.

\section{Matrix Decompositions}

In addition to multiplying matrices, we can decompose them into some combination of unitary matrices and some value that is representative of the matrix itself. These decompositions can be used to transform a seemingly difficult problem into something more easily analyzed.

Schr\"odinger's equation  Eq.~\eqref{schrodinger} can be solved as a linear equation of the form Eq.~\eqref{genlineq}, allowing methods from linear algebra to be used in solving quantum mechanical eigenvalue problems. Many of these methods boil down to finding certain matrix decompositions and leveraging their properties.

\subsection{Eigenvalue Decompositions}

In the case of Schr\"odinger's equation, it is necessary to determine the energy eigenvalues $E_i$ given the operator $\hat H$. This can be accomplished by solving for the characteristic polynomial as
\begin{equation}
\hat H | \psi \rangle = E | \psi \rangle \Rightarrow (\hat H - \mathbb{I} E) | \psi \rangle = \mathbf{0}
\end{equation}
The energy eigenvalues $E_i$ are those which satisfy $\det(\hat H - \mathbb{I} E_i) = 0$, and their corresponding non-zero eigenvectors satisfy $(\hat H - \mathbb{I} E_i) | \psi_i \rangle = \mathbf{0}$.

If we find a way to express $\hat H$ as
\begin{equation}
    \hat H = \hat U \hat \Lambda \hat U^\dagger
\end{equation}
then we have what is called a similarity transformation, where we would say that $\hat \Lambda$ is similar to $\hat H$. Similarity transformations are important because any two matrices which are similar have the same set of eigenvalues.

The diagonal values of $\hat \Lambda$ contain the eigenvalues $E_i$. Actually solving for these values is dependent on the size of the matrix.

\subsection{Schur Decomposition}

Using a computer can lift the difficulty in finding eigenvalues that analytic methods encounter, as we just illustrated in the previous section. One matrix decomposition is the Schur decomposition in which a square matrix is factorized into a unitary matrix $\hat Q$ and an upper triangular matrix $\hat T$ such that
\begin{equation}
    \hat A = \hat Q \hat T \hat Q^{\dagger}
\end{equation}
The eigenvalues can then be read off of the diagonal of $\hat T$ since it is triangular and similar to $\hat A$. It can be shown that every square matrix $\hat A$ has a Schur decomposition \cite{Trefethen-Bau}.

In the case of a Hermitian matrix, the Schur decomposition is equivalent to the eigenvalue decomposition as $\hat T$ will be diagonal. However, the decomposition may give a different matrix $\hat Q$ when compared against an eigenvalue decomposition since the eigenvectors are not unique and can have some arbitrary phase.

\subsection{QR Decomposition}

The QR decomposition provides a method for solving Eq.~\eqref{genlineq} via simple matrix multiplication. It can be intuitively thought of as finding a unitary matrix $\hat Q$ which transforms $\hat A$ into an upper triangular matrix $\hat R$ as described in Ref.~\onlinecite{Watkins}.
\begin{equation}
    \hat Q^{\dagger} \hat A = \hat R
\end{equation}
which is then typically written
\begin{equation}
    \hat A = \hat Q \hat R
\end{equation}
which can be applied to solve Eq.~\eqref{genlineq} as
\begin{align}
    \hat A \mathbf{x} = \mathbf{b}  \\
    \hat Q \hat R \mathbf{x} = \mathbf{b}  \\
    \hat Q \mathbf{w} = \mathbf{b}  \\
    \mathbf{w} = \hat Q^{\dagger} \mathbf{b}
\end{align}
This procedure involves leveraging the facts that $\hat Q$ is unitary so $\hat Q^{\dagger} = \hat Q^{-1}$ and $\hat R$ is upper triangular so a process known as back substitution may be used to find $\mathbf{x}$ from $\hat R \mathbf{x} = \mathbf{w}$ \cite{Watkins}. 

The QR decomposition runs in $O(N^3)$ where matrices $\hat Q$ and $\hat R$ can be found via a Gram-Schmidt process or Givens rotations, see Ref.~\onlinecite{Ford} for more information.

This is not all that the QR decomposition can be used for, however. The QR decomposition is the key step in all forms of the QR algorithm for finding the eigenvalues of a matrix.

\subsection{LU, LDL, and Cholesky Decompositions}

A matrix can be factorized as
\begin{equation}
    \hat A=\hat L\hat U
\end{equation}
where $\hat L$ is a lower triangular matrix (up to a sequence of permutations) and $\hat U$ is an upper triangular matrix. At least one of these should have ones along the diagonal (typically, this is $\hat L$ as chosen by convention). In reality, this is just a description of general Gaussian elimination where $\hat L$ is obtained by multiplying together all elementary row operations to yield a lower-triangular matrix in $\frac{2}{3} N^3$ flops \cite{Golub-vanLoan} \cite{Trefethen-Bau}.

The LU decomposition can be used to solve Eq.~\eqref{genlineq} as
\begin{align}
    \hat A \mathbf{x} = \mathbf{b}  \\
    \hat L \hat U \mathbf{x} = \mathbf{b}  \\
    \hat L \mathbf{w} = \mathbf{b}
\end{align}
Since $\hat L$ and $\hat U$ are both triangular, $\mathbf{w}$ and $\mathbf{x}$ may be found through forward and back substitution in $O(N^2)$ flops.

This procedure may also be applied to find the inverse of a matrix by solving for each $\mathbf{x}_i$ when $\mathbf{b}_i = \mathbf{e}_i$ where $\mathbf{e}_i$ is a standard basis vector. This equivalently translates to solving $\hat A \hat X = \mathbb{I}$ where $\hat X = \hat A^{-1} = [\mathbf{x}_1,\; \mathbf{x}_2, \ldots,\; \mathbf{x}_N]$ and $\mathbb{I} = [\mathbf{e}_1,\; \mathbf{e}_2, \ldots,\; \mathbf{e}_N]$.

A variant of the LU decomposition is the LDL decomposition, which may be applied to obtain the form $\hat A = \hat L \hat D \hat L^{\dagger}$ in $\frac{1}{3} N^3$ flops where $\hat L$ is lower triangular and $\hat D$ is diagonal \cite{Golub-vanLoan}. Note that $\hat D$ is not necessarily similar to $\hat A$, so the values along the diagonal are not the eigenvalues, in general.

Matrix $\hat A$ of size $N \times N$ is Hermitian positive definite if and only if $\hat A$ is Hermitian and $\mathbf{x}^{\dagger} \hat A \mathbf{x} > 0 \in \mathbb{R}$ for all $\mathbf{x} \in \mathbb{C}^{N}$. Matrices of this form are often encountered in physics \cite{Trefethen-Bau}. Matrices which are Hermitian positive definite can be factorized as
\begin{equation}\label{Cholesky}
    \hat A = \hat R^{\dagger} \hat R
\end{equation}
where $\hat R$ is an upper triangular matrix. This type of decomposition is known as the Cholesky factorization \cite{Golub-vanLoan}. 
Cholesky factorization can be done in $\frac{1}{3} N^3$ flops \cite{Trefethen-Bau}.

Matrix $\hat R$ is found by algorithmically applying elementary row operations of Gaussian elimination to $\hat A$ in a process given by Ref.~\onlinecite{Trefethen-Bau}. The first step would result in $\hat A = \hat R_1^{\dagger} \hat A_1 \hat R_1$ where $R_1$ is upper triangular. Then the second step would give $\hat A = \hat R_1^{\dagger} \hat R_2^{\dagger} \hat A_2 \hat R_2 \hat R_1$. This process repeats for $n$ steps until $\hat A_n = \mathbb{I}$ such that we obtain Eq.~\eqref{Cholesky} where $\hat R = \hat R_n  \ldots \hat R_2 \hat R_1$. Essentially, the Cholesky decomposition is a special case of the LU decomposition, which applies when $\hat A$ is Hermitian positive definite.

Eq.~\eqref{genlineq} can be solved with the LDL or Cholesky decomposition by a similar solution strategy as was applied with the LU decomposition, both running in $O(N^2)$ \cite{Golub-vanLoan}.

The LU and LDL decompositions may not be entirely stable and other techniques such as partial pivoting may be required. In contrast, the Cholesky decomposition for Hermitian positive definite matrices is stable \cite{Golub-vanLoan} and so it can be applied to computational problems in physics.

\subsection{Singular Value Decomposition}

One of the most valuable decompositions for generic matrices is the singular value decomposition (SVD). The SVD has recently found popular use as the fundamental tool for the principal component analysis in machine learning \cite{Wang-Zhu} and is a key tool in many tensor network algorithms for quantum systems \cite{bakerCJP21}.

This decomposition works on rectangular matrices, in contrast to the eigenvalue decomposition. We can express the SVD as
\begin{equation}
\hat A=\hat U\hat D\hat V^\dagger
\end{equation}
where $\hat A$ is $M\times N$, $\hat U$ is $M \times M$, $\hat D$ is $M \times N$ with entries (singular values) only along the diagonal, and $\hat V$ is $N \times N$. The diagonal matrix $\hat D$ will have $m=\min(M,N)$ singular values and is often truncated to be $\hat U:\;M\times m$, $\hat D:\; m\times m$, and $\hat V:\; m\times N$ without loss of generality.

We can see that this decomposition can be recovered from the rectangular matrix $\hat A$ by taking the eigenvalue decomposition of $\hat A\hat A^\dagger=\hat U \hat D^2\hat U^\dagger$ and $\hat A^\dagger\hat A=\hat V \hat D^2\hat V^\dagger$. A reasonable question to ask at this point is how stable the eigenvalue decompositions of the square matrices $\hat A\hat A^\dagger$ and $\hat A^\dagger\hat A$ are. In other words, can we use two eigenvalue decompositions to compute the result here? The answer is no for many applications in physics. One would get the elements of the diagonal matrix $\hat D^2$ but trying to take the square root would cause a numerical imprecision. For many computations on quantum systems, the error incurred is not acceptable.

The algorithm to compute the SVD is somewhat complicated and can be found in Ref.~\onlinecite{Golub-vanLoan}, but the general idea is to create the super matrix
\begin{equation}
\check A=\left(\begin{array}{cc}
\hat 0^{M,M} & \hat A^\dagger\\
\hat A & \hat 0^{N,N}
\end{array}\right)
\end{equation}
where $\hat 0^{k,k}$ is a matrix of zeros of dimension $k\times k$. The eigenvalue decomposition of the super matrix is then taken as
\begin{equation}
\check A=\check U\check \Lambda\check U^\dagger
\end{equation}
where $\check U$ is a unitary $(M+N) \times (M+N)$ matrix such that the column vectors of $\hat U$ and $\hat V$ come from the columns of $\check U$. Each positive eigenvalue of $\check \Lambda$ is taken to be a singular value of $\hat D$ and the corresponding column of $\check U$ has rows 1 through $a$ taken to form a column vector of $\hat V$ and rows $a+1$ through $a+b$ used to form a column vector of $\hat U$. Any eigenvalues which are not greater than zero are discarded along with their corresponding column vectors.

This process may result in rectangular matrices for $\hat U$ and $\hat V$ since many vectors are discarded, and $\hat D$ will be square. Multiplying $\hat U \hat D \hat V^{\dagger}$ will result in $\hat A$ correctly, but the unitary matrices will not be complete. To remedy these issues, the incomplete matrices $\hat U$ and $\hat V$ can each be given as input to the QR decomposition, with the resulting $\hat Q$ matrices being the complete unitary matrices to replace $\hat U$ and $\hat V$, respectively; and $\hat D$ must be padded with zeros to have it be of size $M \times N$. Note that the signs of column vectors in $\hat Q$ may end up swapped as some implementations of the QR decomposition are sign agnostic, in which case another step to swap the signs back is required.

Actually formulating the problem and writing out the super matrix is very memory inefficient. Typically, a linear algebra routine only calls the relevant blocks of the super matrix and does not store the whole thing.

\section{Determination of eigenvalues}

The first hurdle in any quantum application is to model the problem correctly. Once that model is obtained (some Hamiltonian for example) one must face the arduous task of actually solving it. For some problems, nice properties such at symmetries and boundary conditions may be leveraged to solve them analytically by hand. For most however, this is not the case.

To solve a quantum system means to find its eigenstates relative to some selected basis. These types of problems can be solved using stable algorithms which apply methods from linear algebra. The key considerations that go into these algorithms are minimizing time-complexity and error.

The eigenvalue problem can be solved analytically for some small cases. However, as matrices increase in size, these simple methods will not suffice. Computational power becomes a requirement. However, the analytic solutions for small matrices can still be useful to know in forming an understanding of the problem at hand.

Consider the case of the $2 \times 2$ matrix, 
 \begin{equation} \label{2by2}
 \hat A=\left(\begin{array}{cc}
 a & b\\
 c & d
 \end{array}\right)
 \end{equation}
the solution is determined by the determinant of $\det(\hat A-\lambda \mathbb{I})$
\begin{equation}
\det\left(\begin{array}{cc}
a - \lambda & b\\
c & d - \lambda
\end{array}\right)=(a-\lambda)(d-\lambda)-cb\overset!=0
\end{equation}
 where $\lambda$ are the eigenvalues, $E_i$. This is quadratic in $\lambda$ and has solutions \cite{boas2006mathematical}
\begin{equation}
\lambda=\frac{(a+d)\pm\sqrt{(a+d)^2-4(ad-cb)}}{2}
\end{equation}

There also exists a general equation for the eigenvalues of $3 \times 3$ matrices obtained by solving $\det(A - \lambda \mathbb{I})=0$ as a cubic equation for a general matrix $A$. The same can be said for $4 \times 4$ matrices with a quartic equation. The expressions for both the $3\times3$ and the $4\times 4$ case are very long to write out in full. 

But since there exist no general algebraic formulas to quintic equations and those of higher order, there exist no general equations for the eigenvalues of $5 \times 5$ matrices and larger. Therefore, past these small matrices, more advanced methods are required.

One could imagine that a root finding algorithm could be used, such as the Newton-Raphson method. This can work to some degree for relatively small matrices; however, it is not ideal as it can lead to the propagation of numerical errors. Round-off errors can be mitigated with a process known as root polishing \cite{Chapra-Canale}, but to a limited degree when working polynomials of many terms.

As we will see, a two-step process can sometimes be applied where a matrix is converted into a favorable form by one algorithm, then solved by another in order to reduce the total number of operations. To minimize error, we must ensure that both of these steps are stable and that they do not run for too many iterations. It must be kept in mind that each iteration of a given algorithm accumulates compounding error due to approximations and rounding. The error incurred at each step depends on the operation performed and the input given. For example, giving an ill-conditioned matrix as input may limit the number of iterations which can be performed as errors often compound aggressively.

If a particular algorithm is not stable, it can often be improved through a series of transformations or by other means.

\subsection{Hessenberg Forms and Householder Transformations}\label{HessenbergTransform}

It would be highly convenient if $\hat A$ provided was of upper Hessenberg form
\begin{equation}\label{Hessenberg}
    \hat A = \left(\begin{array}{ccccccc}
    a_{11} & a_{12} &a_{13} & \cdots & a_{1,N-1} & a_{1,N} \\
    a_{21} & a_{22} & a_{23} &\cdots & a_{2,N-1} & a_{2,N}\\
    0 & a_{32} & a_{33} & \ddots & a_{3,N-1} & a_{3,N} \\
    0 & 0 & a_{43} & \ddots & a_{4,N-1} & a_{4,N} \\
    0 & 0 & 0 & \ddots & \vdots& \vdots \\
    \vdots & \vdots& \ddots & \ddots & a_{N-1,N-1} & a_{N-1,N}\\
    0 & 0 & \cdots  & 0 & a_{N,N-1} & a_{N,N}
\end{array}\right)
\end{equation}
where the triangular portion below the first subdiagonal of the matrix contains only zeros.

One of the key advantages of using matrices in Hessenberg form is their favorable running time. In Code Block~\ref{code:hessenberg} the total number of operations required for  multiplication between two upper Hessenberg matrices is greatly reduced by leveraging the matrix structure. Reduction in the total number of operations is a common theme for algorithms which utilize Hessenberg matrices, as we will see with the Hessenberg QR algorithm.

For quantum problems, an input Hamiltonian $\hat H$ will typically not be in Hessenberg form. In general, it is very uncommon to have an input matrix given in this form. Based on this scarcity in quantum problems, it is reasonable therefore to ask how costly it is to convert a matrix into Hessenberg form.

To convert any matrix into Hessenberg form, Householder transformations are used to generate a unitary transformation which leave only a set number of elements in a selected column as non-zero. The general form that the unitary Householder transformation will take is given according to Ref.~\onlinecite{Ford} as
\begin{equation}\label{equation:householder}
\hat U = \mathbb{I} - \frac{2\mathbf{u} \mathbf{u}^\dagger}{\mathbf{u}^\dagger \mathbf{u}}
\end{equation}
The transformation $\hat U$ (and thereby vector $\mathbf{u }$) must be constructed to ensure that all elements of the $k$th column with index less than or equal to $k+1$ of $\hat A$ are non-zero and all elements of index greater than $k+1$ are zero upon application $\hat U \hat A$. In other words, taking $\mathbf{a_k}$ as the $k$th column of $\hat A$, we must construct and apply transformation $\hat U_k$ according to Eq.~\eqref{equation:householder} such that $\hat U_k \mathbf{a_k} = (\text{b}_{1},\;\text{b}_{2},\ldots,\;\text{b}_{k},\;\text{b}_{k+1},\;0,\;0,\ldots,\;0)^T$.

Ref.~\onlinecite{Garcia-Horn} gives the basic procedure to convert a matrix into Hessenberg form, which has been modified slightly for pedagogical purposes in the following:

\begin{algorithm}[H]\label{alg:hessenberg}
    \caption{Transformation to Hessenberg form}
    \label{RQBL}
    \begin{algorithmic}[1]
        \State Let $\hat A_0 = \hat A$
        \For{$k$ in $1$ to $N-2$}
        \State Compute $\hat U_k$
        \State $\hat A_k = \hat U_k \hat A_{k-1} \hat U_k^{\dagger}$
        \EndFor
        \State return $\hat A_{N-2}$
    \end{algorithmic}
\end{algorithm}

On each step, $\hat U_k$ is computed by first obtaining a vector $\mathbf{v}_k$ as
\begin{equation}
    \mathbf{v}_k =
    \begin{cases}
        |\mathbf{a}_k| \mathbf{e}_k - \mathbf{a}_k & a_{k,1} = 0\\
        |\mathbf{a}_k| \mathbf{e}_k - \frac{\overline{a_{k,1}}}{|a_{k,1}|}\mathbf{a}_k & a_{k,1} \neq 0\\
    \end{cases}
\end{equation}
where $\mathbf{a}_k$ is the $k$th column vector of $\hat A_k$ with the first $k$ row elements omitted such that it has $N-k$ total elements, and $\mathbf{e}_k$ is the unit vector with $N-k$ elements of the form $\mathbf{e}_k = (1,\; 0,\; 0,\ldots,\; 0)^T$. Matrix $\hat V_k^{(N-k) \times (N-k)}$ can be formed via Eq.~\eqref{equation:householder} taking $\mathbf{u}$ as $\mathbf{v}_k$. A Householder transformation with the desired properties may then be created by performing a direct sum between a $k \times k$ identity matrix and $\hat V_k^{(N-k) \times (N-k)}$.
\begin{equation}
    \hat U_k = \hat{\mathbb{I}}^{k \times k} \bigoplus \hat V_k^{N-k \times N-k}
\end{equation}

$\hat U_k$ is then applied to obtain
\begin{equation}\label{householdertransform}
    \hat A_{k+1} = \hat U_k \hat A_k \hat U_k^{\dagger}
\end{equation}
and after all iterations are performed, the resulting matrix $\hat A_{N-2}$ will be in Hessenberg form.

If $\hat A$ is Hermitian, then applying the transformation of Eq.~\eqref{householdertransform} will result in $\hat A_{k+1}$ also being Hermitian \cite{Eidelman}. Therefore, a Hermitian matrix transformed into Hessenberg form by this process will be tridiagonal.

Code Block~\ref{code:householder} displays the steps for transforming any matrix into a Hessenberg form matrix. This similarity transformation preserves the eigenvalues of $\hat A$.

\begin{lstlisting}[firstline=1, lastline=40, firstnumber=1, tabsize=2, language=Julia, caption={Householder transformation}, label={code:householder}]{julia}
using LinearAlgebra
import TensorPACK

function householder(A, k)
    n = size(A,2)
    Id = Matrix{eltype(A)}(I, k, k)
    e = zeros(n-k)
    e[1] = 1
    v = norm(A[k+1:end,k]) * e 
    if A[k+1,k] == 0
        v -= A[k+1:end,k]
    else
        normA = conj(A[k+1,k])
        normA /= abs(A[k+1,k])
        v -= A[k+1:end,k] * normA
    end
    sU = Matrix{eltype(A)}(I, n-k, n-k) 
    sU -= 2*v*v' / norm(v)^2
    U_k = TensorPACK.directsum(Id, sU)
    return U_k
end

function hessenberg_convert(A)
    for k in 1:size(A,2)-2
        U_k = householder(A, k)
        A = U_k*A*U_k'
    end
    return A
end
\end{lstlisting}

The implementation here is somewhat naive and unoptimized, but it can be shown that reduction to Hessenberg form via more complex techniques can be done in $\frac{10}{3}N^3$ operations \cite{Ford}. Implementations of the transformation algorithm are very stable and produce reliable results for any matrix \cite{Ford}.

Only square matrices were considered here, though the Householder transformation can be applied to rectangular matrices as well by computing a different unitary transformation. The rectangular Householder transformation is crucial in efficient implementations of the singular value decomposition.

\subsection{The QR Algorithm and its Variants}\label{QRalg}

The QR algorithm gains its name from the application of the QR decomposition done at each step. This decomposition can be used to solve Eq.~\eqref{genlineq}. The QR algorithm is a powerful tool for computational quantum mechanics as it can be used to find all the eigenvalues of a given matrix. The most basic form of the QR algorithm is quite straight forward, but additional complexity is required to improve stability.

\subsubsection{Unshifted QR Algorithm}

The basic unshifted QR algorithm is best encapsulated by Ref.~\onlinecite{Ford} in the following steps:

\begin{algorithm}[H]\label{alg:QR}
    \caption{Unshifted QR method}
    \label{RQBL}
    \begin{algorithmic}[1]
        \State Initialize $\hat A_1 = \hat A$
        \While{$A_k$ is not in (approximate) upper triangular form}
        \State Perform the decomposition $\hat A_k = \hat Q_k \hat R_k$ 
        \State Take $\hat A_{k+1} = \hat R_k \hat Q_k$
        \EndWhile
        \State return the final $\hat A_{k}$
    \end{algorithmic}
\end{algorithm}

The essential idea of this algorithm is that on the first iteration we find the QR decomposition of $\hat A$ as $\hat R_1 = \hat Q_1^{\dagger} \hat A_1$ such that $\hat A_2 = \hat R_1 \hat Q_1 = \hat Q_1^{\dagger} \hat A_1 \hat Q_1$. Since $\hat Q_1$ is a unitary matrix, this is a similarity transformation which preserves the eigenvalues of $\hat A_1$ upon transformation to $\hat A_2$. This argument holds at each step. Upon repeated iteration, $\hat A_k$ converges to an upper triangular matrix and so the eigenvalues may be read directly from the diagonal.

Note that if the input matrix is Hermitian, then each of the similarity transformations will preserve the Hermitian property and so $\hat A_k$ will converge to a diagonal matrix. This gives a method for the diagonalization of Hermitian matrices. Each of the unitary matrices, $\hat Q_i$, can be used to obtain the eigenvectors in a stable way through repeated matrix multiplication. The columns of $\hat U = \hat Q_1 \hat Q_2 \ldots \hat Q_k$ then give the eigenvectors of the Hermitian input matrix $\hat A$.

Numerical errors must be balanced. Applying too few iterations will result in a matrix which is not quite triangular and so the calculated eigenvalue spectrum will be incorrect. However, applying too many iterations may cause floating point errors to accumulate and can also result in an incorrect set of eigenvalues. Thus, there is an optimal number of iterations. Though this optimal number of iterations may depend on the conditioning of the input matrix as this can determine how much error is incurred at each step.

In general, the algorithm should be run until the lower off-diagonal elements have a sufficiently small norm. For time-complexity, it is assumed that the number of required iterations scales linearly with matrix size. Each QR decomposition and matrix multiplication requires $O(N^3)$ flops, giving the QR algorithm a time-complexity of $O(N^4)$ \cite{Ford}.

\subsubsection{Utilizing Hessenberg Form}

This brings us to why so much attention has been focussed on Hessenberg form matrices. It can be shown that finding the QR decomposition of a Hessenberg matrix can be done in $O(N^2)$ rather than $O(N^3)$ \cite{Ford}. Additionally, in Hessenberg form fewer iterations of the QR algorithm are required. This makes intuitive sense as Hessenberg matrices are nearly in triangular form already.

Recall, that conversion to Hessenberg form for a general matrix costs $O(N^3)$ operations. Then the QR algorithm is considered to run in $O(N^2)$, giving an overall time-complexity of $O(N^3)$ rather than $O(N^4)$ \cite{Ford}.

However, this comes with a caveat. The rate of convergence of the QR algorithm for Hessenberg matrices decreases as the difference in the magnitudes of the smallest eigenvalues decreases \cite{Ford}. This motivates finding other strategies to improve the rate of convergence as it can be impractical in some cases.

\subsubsection{Shifted QR Algorithm}

The rate of convergence for the QR algorithm when applied to a Hessenberg matrix may be improved by incorporating a shift at each step. Given matrix $\hat A$ in Hessenberg form, the steps of the shifted QR algorithm are as follows \cite{Golub-vanLoan}:

\begin{algorithm}[H]\label{alg:shiftedQR}
    \caption{Shifted QR method}
    \label{RQBL}
    \begin{algorithmic}[1]
        \State Initialize $\hat A_1 = \hat A$
        \While{$A_k$ is not in (approximate) upper triangular form}
        \State Perform the shifted decomposition $\hat A_k - \mu_k \mathbb{I} = \hat Q_k \hat R_k$
        \State Take $\hat A_{k+1} = \hat R_k \hat Q_k + \mu_k \mathbb{I}$
        \EndWhile
        \State return the final $\hat A_{k}$
    \end{algorithmic}
\end{algorithm}

The key idea here is to choose $\mu_k$ as an approximate eigenvalue of $\hat A$. There are many ways in which $\mu_k$ may be chosen, but the simplest and most intuitive is to take $\mu_k$ as a diagonal element of $\hat A$ at each step, a strategy called the Rayleigh quotient shift \cite{Ford}. Since $\hat A$ is in Hessenberg form, $a_{j,j}$ is a decent approximation. Typically, the last diagonal element $\mu_k = a_{N,N}$ will be taken as the shift parameter until $a_{N-1,N}\approx0$ after a sufficient number of iterations, then it will be taken as $\mu_k = a_{N-1,N-1}$ until $a_{N-2,N-1}\approx0$, and so on \cite{Ford}. The process is complete once the matrix is in triangular form.

For other choices of shift parameter $\mu_k$ see Ref.~\onlinecite{Golub-vanLoan}.

\subsubsection{Double Shifted QR Algorithm}

In the case of complex numbers along the diagonal of the Hessenberg form matrix, the single shift algorithm will not converge.  It is therefore necessary to define an extended algorithm capable of handling complex values. This can be done by introducing a second shift by $\sigma_k \mathbb{I}$ at each step.

The double shifted QR algorithm takes a Hessenberg form matrix as input and performs two shifts in the iterative step \cite{Golub-vanLoan}.

\begin{algorithm}[H]\label{alg:shiftedQR}
    \caption{Unshifted QR method}
    \label{RQBL}
    \begin{algorithmic}[1]
        \State Initialize $\hat A_1 = \hat A$
        \While{$A_k$ is not in (approximate) upper triangular form}
        \State Perform the shifted decomposition $\hat A_k - \mu_k \mathbb{I} = \hat Q_k \hat R_k$
        \State Take $\hat A_k' = \hat R_k \hat Q_k + \mu_k \mathbb{I}$
        \State Perform the shifted decomposition $\hat A_k' - \sigma_k \mathbb{I} = \hat Q_k' \hat R_k'$
        \State Take $\hat A_{k+1} = \hat R_k' \hat Q_k' + \sigma_k \mathbb{I}$
        \EndWhile
        \State return the final $\hat A_{k}$
    \end{algorithmic}
\end{algorithm}

This is the essential idea of the double shifted QR algorithm. However, conversion from $\hat A_k$ to $\hat A_{k+1}$ can be done in $O(N^2)$ operations by specially constructing and applying a Householder transformation in what is called a double implicit shifted QR algorithm \cite{Golub-vanLoan}. It can be shown that the particular Householder transformation constructed yields the same resulting matrix as the two shifts would when applied. Modern methods use implicit shifts for efficiency.

For an example of an implicit double shift QR algorithm see the Francis QR step in Ref.~\onlinecite{Golub-vanLoan}. For an example of a modern QR technique, see Refs.~\onlinecite{Braman-1, Braman-2}.

\subsection{Power Method}\label{powermethod}

The power method is a linear algebra algorithm used to determine the largest eigenvalue of a matrix and its associated eigenvector. The algorithm depends on the fact that the exponential of the eigenvalue of greatest magnitude will grow much faster than the second-greatest eigenvalue, and thus all other eigenvalues. An operator $\hat A$ with the greatest eigenvalue $E_x$ taken to the $n$th power, for sufficiently large $n$, will have its eigenvalue spectrum dominated by $E_x^n$.

Using the notation in quantum mechanics, we can note that the application of an operator onto a general wavefunction, $\Psi$, is best thought of when $\Psi$ is expanded in terms of eigenfunctions of the operators, $\psi_k$. So, to rewrite $\Psi$, we obtain
\begin{equation}
|\Psi\rangle=\sum_k c_k|\psi_k\rangle
\end{equation}
where the coefficients $c_k$ are some complex number denoting the inner product $\langle\psi_k|\Psi\rangle=c_k$.

Similarly, by the above statements, we have the operator $\hat A$ expanded as
\begin{equation}
\hat A=\sum_kE_k|\psi_k\rangle\langle\psi_k|
\end{equation}
where $E_k$ is an eigenvalue. If we apply $\hat A$ onto $\Psi$ for a total of $n$ times, then we obtain
\begin{equation}
\hat A^n|\Psi\rangle=\sum_kc_kE_k^n|\psi_k\rangle
\end{equation}
Even if $c_x \ll c_k$, we will have $E_x^n \ggg E_k^n\; \forall\; k \neq x$ after enough iterations. If there is a gap ({\it i.e.}, a finite difference between adjacent energy eigenvalues), then it will grow exponentially with successive iterations of $\hat A$, no matter how small. This means that repeatedly applying the matrix $\hat A^n$ exposes the most extremal eigensolution.

In practice, normalization must occur to correct any numerical errors. Upon each application of $\hat A$, $\Psi$ is normalized to be of unit length. The algorithm may converge slowly, but it is best optimized for sparse matrices.

\subsection{The Lanczos Method}\label{lanczos}

Hamiltonians are often formulated as sparse matrices. These matrices may be converted into Hessenberg form and solved using the QR algorithm. Matrix size scales exponentially with the number of particles involved, so the QR algorithm eventually reaches a limit in its applicability. It is desirable to take advantage of the sparsity in order to solve the eigenvalue problem in fewer operations. This is where the Lanczos method comes in. Some applications of these types of methods can be applied in a matrix-less way for matrices with a specific structure (i.e.~methods for matrices of a certain structure in which the underlying matrix need not be stored or explicitly formed)~\cite{bogoya2024fast} or with extrapolation techniques in power methods for large problems~\cite{serra2005jordan,brezinski2006pagerank}, but the focus here will be when there is an input matrix.

The Lanczos method is useful when extremal solutions are desired rather than all solutions, though it can still be applied to find all solutions \cite{Golub-vanLoan}. For applications to statistical mechanics, a few of the lowest energy states of a system can often suffice.

The essential idea of the Lanczos method is to compute an orthonormal basis for the Krylov subspace
\begin{align}
    K_k(\hat A, \mathbf{q}_1) &= \mathrm{span}\{\mathbf{q}_1, \hat A \mathbf{q}_1, \hat A^2 \mathbf{q}_1, \ldots, \hat A^{k-1} \mathbf{q}_1\} \nonumber \\
        &= \mathrm{span}\{\mathbf{q}_1, \mathbf{q}_2, \mathbf{q}_3, \ldots, \mathbf{q}_k\}
\end{align}
via tridiagonalization of $\hat A$ as
\begin{equation}
    \hat T_k = \hat Q_k^{\dagger} \hat A \hat Q_k
\end{equation}
where $\hat Q_k = [\mathbf{q}_1, \mathbf{q}_2, \ldots, \mathbf{q}_k]$ is the matrix of orthonormal Lanczos vectors and $\mathbf{q}_1$ is some initial Lanczos vector of unit length \cite{Golub-vanLoan}. $\hat T_k$ can then be used to find $k$ of the most extremal eigensolutions of $\hat A$ by applying the QR algorithm or the divide and conquer algorithm of Sec.~\ref{daq}.

$\hat T_k$ takes the form
\begin{equation}
    \hat T_k = \left(\begin{array}{ccccccc}
    \alpha_1 & \beta_1 & 0 & 0 & \cdots & 0\\
    \beta_1 & \alpha_2 & \beta_2 & 0 & \cdots & 0\\
    0 & \beta_2 & \alpha_3 & \beta_3 & \cdots & 0\\
    0 & 0 & \beta_3 & \alpha_4 &\ddots & \vdots\\
    \vdots & \vdots & \vdots & \ddots & \ddots & \beta_{k-1}\\
    0 & 0 & 0 & \cdots & \beta_{k-1} & \alpha_k
    \end{array}\right)
\end{equation}
where $\alpha_i$ and $\beta_i$ are the Lanczos coefficients computed on the $i$th step of the algorithm. The column vectors of $\hat Q$ are guaranteed to be orthogonal to each other by construction. On each step, the Lanczos coefficients are calculated as
\begin{equation}
\alpha_i = \mathbf{q}_i^{\dagger} \hat A \mathbf{q}_i
\end{equation}
and
\begin{equation}
\beta_i = ||(\hat A - \alpha_i \mathbb{I})\mathbf{q}_i - \beta_{i-1}\mathbf{q}_{i-1}||_2
\end{equation}
where
\begin{equation}
    \mathbf{q}_{i+1} = \frac{ (\hat A - \alpha_i \mathbb{I})\mathbf{q}_i - \beta_{i-1}\mathbf{q}_{i-1} }{ \beta_i }
\end{equation}

The Lanczos algorithm can be applied to obtain $\hat T_k$ given some $\hat A$ and arbitrary unit vector $\mathbf{q}_1$ in just $k$ iterations, requiring only two storage vectors. The basic algorithm is prone to a significant accumulation of errors resulting in a loss of orthogonality, though methods exist to reduce error \cite{Golub-vanLoan}.

In Ref.~\onlinecite{Golub-vanLoan}, Golub and Van Loan  give a succinct description of the algorithm which we borrow from to produce Code Block~\ref{code:Lanczos}, although our implementation is slightly less efficient in its memory consumption for pedagogical purposes.

\begin{lstlisting}[tabsize=2, firstnumber=1,language=Julia, caption={Lanczos method}, label={code:Lanczos}]{julia}
using LinearAlgebra

function Lanczos(A, q, k)
	n = size(A, 1)
	v = zeros(eltype(q), n)
	T = zeros(eltype(A), k, k)
	beta = 1

	for i in 1:k
		if i != 1
			for j in 1:n
				t = q[j]
				q[j] = v[j] / beta
				v[j] = -beta * t
			end
		end

		v = v + A * q
		alpha = q' * v
		v = v - alpha * q
		beta = norm(v)

		T[i,i] = alpha
		if i < k
			T[i, i+1] = beta
			T[i+1, i] = beta
		end
	end
	return T
end
\end{lstlisting}

Implementations of the Lanczos algorithm run in approximately $O(kpN)$ flops where $p$ is the average number of non-zero entries in each row of $\hat A$ since only matrix-vector multiplications are required on each iteration \cite{Golub-vanLoan}. The key to its efficiency is that orthogonality is guaranteed by construction and on each step the most expensive operation is a matrix-vector multiplication, running in $O(pN)$. 

The approximate eigenvectors may be obtained along with their corresponding eigenvalues when $\hat T$ is diagonalized via the QR algorithm as $\hat T = \hat U \hat D \hat U^{\dagger}$. Application of the QR algorithm to a tridiagonal matrix to obtain the eigenvalues runs in $O(N^2)$.

\subsection{Divide and Conquer}\label{daq}

In comparison to variants of the QR algorithm, the divide and conquer algorithm is more efficient when finding both the eigenvalues and eigenvectors of sufficiently large matrices. This is the algorithm that is used for eigenvalue decompositions in many high-level codes by default. One especially large advantage for quantum applications is that it is parallelizable, meaning that a large system can be broken down into subproblems which may then be solved in parallel.

The divide and conquer algorithm requires that the input matrix be in Hermitian tridiagonal form, which is guaranteed for any Hermitian matrix transformed into Hessenberg form or if $\hat T$ is found via the Lanczos algorithm. 

The goal is to find the decomposition 
\begin{equation}
    \hat Q^{\dagger} \hat T \hat Q = \hat \Lambda,
\end{equation}
where $\hat \Lambda$ is a diagonal matrix of eigenvalues and $\hat Q$ is unitary. This is achieved by expressing the $N \times N$ symmetric tridiagonal matrix $\hat T$ as
\begin{equation}\label{divide}
    \hat T = \left(\begin{array}{c|c}
    \hat T_1 & \hat 0^{m \times (N-m)} \\
    \hline
    \hat 0^{(N-m) \times m} & \hat T_2 \\
    \end{array}\right) + \rho \mathbf{v} \mathbf{v}^{\dagger}
\end{equation}
where $\hat T_1$ and $\hat T_2$ are defined as
\begin{equation}
    \hat T_1 = \left(\begin{array}{ccccccc}
        \alpha_1 & \beta_1 & 0 & 0 & \cdots & 0\\
        \beta_1 & \alpha_2 & \beta_2 & 0 & \cdots & 0\\
        0 & \beta_2 & \alpha_3 & \beta_3 & \cdots & 0\\
        0 & 0 & \beta_3 & \alpha_4 &\ddots & \vdots\\
        \vdots & \vdots & \vdots & \ddots & \ddots & \beta_{m-1}\\
        0 & 0 & 0 & \cdots & \beta_{m-1} & \gamma_m
    \end{array}\right)
\end{equation}
and
\begin{equation}
    \hat T_2 = \left(\begin{array}{ccccccc}
        \gamma_{m+1} & \beta_{m+1} & 0 & 0 & \cdots & 0\\
        \beta_{m+1} & \alpha_{m+2} & \beta_{m+2} & 0 & \cdots & 0\\
        0 & \beta_{m+2} & \alpha_{m+3} & \beta_{m+3} & \cdots & 0\\
        0 & 0 & \beta_{m+3} & \alpha_{m+4} &\ddots & \vdots\\
        \vdots & \vdots & \vdots & \ddots & \ddots & \beta_{N-1}\\
        0 & 0 & 0 & \cdots & \beta_{N-1} & \alpha_N
    \end{array}\right)
\end{equation}
with $\mathbf{v} = (0, \ldots, 0, 1, \theta, 0, \ldots, 0)^T$ such that only the $m$th and $(m+1)$th entries are non-zero, $\rho \theta = \beta_m$, $\gamma_m = \alpha_m - \rho$, and $\gamma_{m+1} = \alpha_{m+1} - \rho \theta^2$ \cite{Golub-vanLoan}.

To write this all out explicitly, we have
\begin{align}
    \hat T = & \left(\begin{array}{c|c}
    \hat T_1 & \hat 0^{m \times (N-m)} \\
    \hline
    0^{(m-N) \times m} & \hat T_2 \\
    \end{array}\right) +\\
    & \left(\begin{array}{c|cc|c}
        & 0 & 0 & \\
        \hat 0^{(m-1) \times (m-1)} & \vdots & \vdots & \hat 0^{(m-1) \times (N-m-1)} \\
        & 0 & 0 & \\
        \hline
        0 \ldots 0 & \rho & \theta \rho & 0 \ldots 0\\
        0 \ldots 0 & \theta \rho & \theta^2 \rho & 0 \ldots 0 \\
        \hline
        & 0 & 0 & \\
        \hat 0^{(N-m-1) \times (m-1)} & \vdots & \vdots & \hat 0^{(N-m-1) \times (N-m-1)} \\
        & 0 & 0 &\\
    \end{array}\right)\nonumber
\end{align}

Both $\hat T_1$ and $\hat T_2$ are then subdivided further by this same process, which can happen at most $\lfloor \log_2N - 1 \rfloor$ times. This description outlines the divide step.

The conquer step begins by finding the eigenvalue decomposition of each tridiagonal submatrix as $\hat Q_i^{\dagger} \hat T_i \hat Q_i = \hat D_i$. Let $\hat Y$ represent the supermatrix formed by two tridiagonal submatrices at some depth according to Eq.~\eqref{divide}, then the following similarity transformation of $\hat Y$ yields
\begin{align}
    \hat \Delta &= \hat U^{\dagger} \hat Y \hat U  \\
    &= \left(\begin{array}{c|c}
        \hat Q_a & \hat 0 \\
        \hline
        \hat 0 & \hat Q_b \\
    \end{array}\right)^{\dagger}
    \left(\left(\begin{array}{c|c}
        \hat T_a & \hat 0 \\
        \hline
        \hat 0 & \hat T_b \\
    \end{array}\right) + \rho \mathbf{v} \mathbf{v}^{\dagger}\right)
    \left(\begin{array}{c|c}
        \hat Q_a & \hat 0 \\
        \hline
        \hat 0 & \hat Q_b \\
    \end{array}\right) \nonumber \\
    &= \left(\begin{array}{c|c}
        \hat D_a & \hat 0 \\
        \hline
        \hat 0 & \hat D_b \\
    \end{array}\right) + \rho \mathbf{z} \mathbf{z}^{\dagger}
\end{align}
where $\mathbf{z} = \hat U^{\dagger} \mathbf{v}$. $\hat \Delta$ is then a diagonal matrix plus a rank-1 correction, the eigenvalue decomposition of which can be found relatively quickly. Our focus remains on the algorithmic procedures rather than mathematical details, so we allege that this can be done, but for more details see Ref.~\onlinecite{Golub-vanLoan}. 

The conquer step may then be run again on $\hat \Delta$ and a neighboring submatrix of the same size. This step is repeatedly run on the matrices produced until the eigenvalue decomposition of $\hat T$ is obtained.

Notice that since each eigenvalue problem is independent of the others for any given depth, all the submatrices of the same depth may have the conquer step applied in parallel.

\subsection{Implementations}

Freely available libraries such as LAPACK, OpenBLAS, Intel MKL, and others have decades of testing and are extremely reliable. In extensions to graphical processing units (GPUs) where linear algebra can be performed in parallel, we note cuSOLVER, cuBLAS, and ROCm as current software options. In addition, many of these libraries also use advanced techniques, further complicating implementation. They should be used instead of programming new routines. While the speed and efficiency can be matched in some cases, it is very unlikely that any improvement can be made on the existing algorithms in a reasonable amount of time. This means that one faces a choice of either reprogramming and improving the existing routines or solving actual physics problems.

In particular, implementing the eigenvalue decomposition in a stable way is very difficult. One needs an efficient shifted-QR algorithm in place to handle a variety of situations. One also has to create memory on the computer that is reused. The algorithms programmed into LAPACK rely on efficient BLAS operations and use an economical amount of memory. These operations include advanced techniques for efficiency like memory pre-fetching, caching, vectorization, etc.

Even though the standard advice is not to reprogram these methods, there are situations where these methods are not efficient. Computations for sparse matrices can sometimes be accelerated by leveraging matrix structure. By specifically crafting algorithms this way, eigenvalue problems for much larger systems can be solved.

\section{Extension to tensors}

So far in this review, we have discussed vectors and matrices. There is a generalization of a notion of a tensor which is defined based on the rank of a tensor. For example, a rank-0 tensor is a scalar value. A rank-1 tensor is a vector. A rank-2 tensor is a matrix.

The main note about higher-rank tensors is that they can be reshaped and permuted to create the matrix equivalent of a tensor. For example, if we are presented with a rank-3 tensor, then we can group two of the indices together as the column of a matrix. The remaining index can be the row. This then converts the tensor contraction and decomposition methods to the matrix methods we covered here. A wealth of algorithms can be formulated based on these constructions.

For applications of tensors in quantum physics, and explicit use of the reshaping and permuting of tensors to matrix equivalents, we point the interested reader to Ref.~\onlinecite{bakerCJP21} which discusses a class of algorithms relying on entanglement renormalization formulated on a tensor network.

The basic functions needed for operations on tensors is contained in a custom built library in Ref.~\onlinecite{tenpack}.

\section{Conclusion}

Computational linear algebra methods are crucial in solving quantum mechanical systems. In this review, we motivated the problem of solving systems of linear equations by introducing the time-independent Schr\"odinger equation, discussed how quantum systems can be formulated in the language of linear algebra, and covered common ways in which matrix decompositions are obtained and applied in modern libraries. The overall focus was on solving eigenvalue problems, as these give solutions to Schr\"odinger's equation. All the methods discussed may be applied across disciplines.

\section{Acknowledgements}

We thank the anonymous referees who helped improve this paper.

A.D. and K.G. acknowledge support from the Undergraduate Student Research Award (USRA) from NSERC. A.D. and K.G. ~acknowledge the NSERC CREATE in Quantum Computing Program, grant number 543245.

A.D.~is grateful for the support of the Jamie Cassel's Undergraduate Research Award (JCURA).

K.G.~acknowledges support from the Valerie Kuehne Undergraduate Research Award (VKURA) and the Summer Emerging Research Award (SERA) from the Faculty of Science at the University of Victoria.

S.H. acknowledges support from the Digital Research Alliance of Canada (alliancecan.ca).

This research was undertaken, in part, thanks to funding from the Canada Research Chairs Program (CRC-2021-00257). This work has been supported in part by the Natural Sciences and Engineering Research Council of Canada (NSERC) under grants RGPIN-2023-05510 and DGECR-2023-00026. 

T.E.B.~is grateful to the US-UK Fulbright Commission for financial support under the Fulbright U.S. Scholarship programme as hosted by the University of York.  This research was undertaken in part thanks to funding from the Bureau of Education and Cultural Affairs from the United States Department of State. T.E.B.~also thanks the availability of the King's Manor library at the University of York, United Kingdom where parts of the notes for this work were first created.


\begin{thebibliography}{30}
\bibitem{anderson1999lapack} Edward Anderson, Zhaojun Bai, Christian Bischof, L Susan Blackford, James Demmel, Jack Dongarra, Jeremy Du Croz, Anne Greenbaum, Sven Hammarling, Alan McKenney, {\it et al., LAPACK users’ guide} (SIAM, 1999).

\bibitem{Berry} Richard Cleve,  Robin Kothari, Dominic W. Berry, Andrew M. Childs and Rolando D. Somma, “Simulating hamiltonian dynamics with a truncated Taylor Series,” Phys. Rev. Lett. 114 (2015), 10.1103/PhysRevLett.114.090502.

\bibitem{townsend2000modern} John S Townsend, {\it A modern approach to quantum mechanics} (University Science Books, 2000).

\bibitem{Boyce} Richard C. Diprima William E. Boyce and Douglas B. Meade, Elementary Differential Equations and Boundary Value Problems - 11th Edition (John Wiley \& Sons, 2017) p. 303.

\bibitem{bakerCJP21} Thomas E Baker, Samuel Desrosiers, Maxime Tremblay, and Martin P Thompson, “Méthodes de calcul avec réseaux de tenseurs en physique,” Canadian Journal of Physics {\bf 99}, 4 (2021); “Basic tensor network computations in physics,” arXiv:1911.11566, p. 20 (2019).

\bibitem{Goodrich} Michael T. Goodrich and Roberto Tamassia {\it Algorithm Design - Foundations, Analysis, and Internet Examples - First Edition} (John Wiley \& Sons, 2002) pp.~13-20

\bibitem{Chapra-Canale} Steven C. Chapra and Raymond P. Canale, {\it Numerical Methods for Engineers - Eighth Edition} (McGraw Hill, 2021) pp. 180–183, 249–274.

\bibitem{Trefethen-Bau} Lloyd N. Trefethen and David Bau III, {\it Numerical Linear Algebra} (Society for Industrial and Applied Mathematics, 1997) pp. 34, 94–95, 172–177, 187–188.

\bibitem{Shafarevich} Igor R. Shafarevich and Alexey O. Remizov {\it Linear Algebra and Geometry}  (Springer Science \& Business Media, 2012) pp.~56-57

\bibitem{Cho} Cho, Jeong-Rae and Cho, Keunhee and Yoon, Hyejin and Rhee, Dong Sop and Lee, Jin Ho, "Application of the Q-Less QR Factorization to Resolve Sparse Linear Over-Constraints" Applied Sciences, {\bf 15}, 13059 (2025)

\bibitem{Kelefouras} Iosif Mporas Vasilios Kelefouras, Angeliki Kritikakou, and Vasilios Kolonias, “A high-performance matrix–matrix multiplication methodology for cpu and gpu architectures,” The Journal of Supercomputing {\bf 72}, 804–844 (2016).

\bibitem{kowarschik2003overview} Markus Kowarschik and Christian Weiß, “An overview of cache optimization techniques and cache-aware numerical algorithms,” Algorithms for memory hierarchies: advanced lectures, 213–232 (2003).

\bibitem{whaley2005minimizing} R Clint Whaley and Antoine Petitet, “Minimizing development and maintenance costs in supporting persistently optimized blas,” Software: Practice and Experience 35, 101–121 (2005).

\bibitem{vanLoan} Bo Kågström, Per Ling, and Charles van Loan, “GEMM-based level 3 BLAS: High-performance model implementations and performance evaluation benchmark,” ACM Trans. Math. Softw. 24, 268–302 (1998).

\bibitem{Geijn} Robert van de Geijn and Kazushige Goto, BLAS (Basic Linear Algebra Subprograms), edited by David Padua (Springer US, 2011) pp. 157–164.

\bibitem{Ascher} Uri M. Ascher and Chen Greif, {\it A First Course on Numerical Methods} (Society for Industrial and Applied Mathematics, 2011) pp. 117–121.



\bibitem{chan1996conjugate} Chan, Raymond H.; Ng, Michael K. Conjugate gradient methods for Toeplitz systems. SIAM Rev. 38 (1996), no. 3, 427–482.

\bibitem{ng2004iterative} Ng, Michael K. Iterative methods for Toeplitz systems. {\it Numerical Mathematics and Scientific Computation.} (Oxford University Press, New York, 2004)

\bibitem{garoni2017generalized} Garoni, Carlo; Serra-Capizzano, Stefano. {\it Generalized locally Toeplitz sequences: theory and applications. Vol. I.} (Springer, Cham, 2017)

\bibitem{garoni2018generalized} Garoni, Carlo; Serra-Capizzano, Stefano. {\it Generalized locally Toeplitz sequences: theory and applications. Vol. II.} (Springer, Cham, 2018)


\bibitem{Golub-vanLoan} Gene H. Golub and Charles F. van Loan, Matrix Computations - Third Edition (The Johns Hopkins University Press, 1996) pp. 16, 80–82, 89–98, 135–143, 223–233, 352–361, 442–449, 470–488.

\bibitem{Hogben} Leslie Hogben, {\it Elementary Linear Algebra - First Edition} (West Publishing Company, 1987) p. 130.

\bibitem{Garcia-Horn} Stephan Ramon Garcia and Roger A. Horn, {\it Matrix Mathematics - A Second Course in Linear Algebra} (Cambridge University Press, 2017).

\bibitem{Hall} Brian C. Hall, {\it Lie Groups, Lie Algebras, and Representations - An Elementary Introduction - Second Edition} (Springer US, 2015) p. 27.

\bibitem{Anton-Rorres} Howard Anton and Chris Rorres, {\it Elementary Linear Algebra - Applications version - 11th Edition} (John Wiley \& Sons, 2014) pp. 30–31, 237–238.

\bibitem{Huang} Greg Henry Jianyu Huang, Tyler Smith and Robert van de Geijn, “Strassen’s algorithm reloaded,” (IEEE Press, 2016) pp. 690–701.

\bibitem{Saad} Yousef Saad, {\it Iterative Methods for Sparse Linear Systems - Second Edition} (Society for Industrial and Applied Mathematics, 2003) pp. 92–95.

\bibitem{Watkins} David S. Watkins, {\it Fundamentals of Matrix Computations- Second Edition} (John Wiley \& Sons, 2002) pp. 190–191.

\bibitem{Ford} William Ford, {\it Numerical Linear Algebra with Applications - Using MATLAB} (Academic Press, 2015) Chap. 14, 17, 18.

\bibitem{Wang-Zhu} Yongchang Wang and Ligu Zhu, “Research and implementation of SVD in machine learning,” (IEEE Press, 2017) pp. 471–475.

\bibitem{boas2006mathematical} Mary L Boas, {\it Mathematical methods in the physical sciences} (Wiley, 2006).

\bibitem{Eidelman} Israel Gohberg Yuli Eidelman and Luca Gemignani, “On the fast reduction of a quasi-separable matrix to Hessenberg and tridiagonal forms,” {\it Linear Algebra and its Applications} 420, 86–101 (2007).

\bibitem{Braman-1} Ralph Byers Karen Braman and Roy Mathias, “The multishift QR algorithm. part i: Maintaining well-focused shifts and level 3 performance,” { SIAM Journal on Matrix Analysis and Applications} {\bf 23}, 929–947 (2002).

\bibitem{Braman-2} Ralph Byers Karen Braman and Roy Mathias, “The multishift QR algorithm. part ii: Aggressive early deflation,” { SIAM Journal on Matrix Analysis and Applications} {\bf 23}, 948–973 (2002).

\bibitem{bogoya2024fast} Bogoya, Manuel; Grudsky, Sergei M.; Serra-Capizzano, Stefano {\it Fast non-Hermitian Toeplitz eigenvalue computations, joining matrixless algorithms and FDE approximation matrices.} { SIAM J. Matrix Anal. Appl.} {\bf 45} (2024), no. 1, 284–305.

\bibitem{serra2005jordan} Serra-Capizzano, Stefano. {\it Jordan canonical form of the Google matrix: a potential contribution to the PageRank computation.} { SIAM J. Matrix Anal. Appl. } {\bf 27} (2005), no. 2, 305–312.

\bibitem{brezinski2006pagerank} Brezinski, Claude; Redivo-Zaglia, Michela "The PageRank vector: properties, computation, approximation, and acceleration". { SIAM J. Matrix Anal. Appl.} {\bf 28} (2006), no. 2, 551–575.

\bibitem{tenpack} Thomas E. Baker, “TENPACK,” https://github.com/bakerte/TensorPACK.jl (2020).

\end{thebibliography}
\end{document}